\def\papertitle{Gauss Circle Lattices with Geometric Convolutions for Synthesizing High Dimensional Image-Source Room Impulse Responses}
\def\paperauthorA{Yuancheng Luo}

\documentclass[twoside,a4paper]{article}
\usepackage{etoolbox}

\usepackage[print]{dafx26v3}

\usepackage{amsmath,amssymb,amsfonts,amsthm}
\usepackage{siunitx}
\usepackage{euscript}
\usepackage[T1]{fontenc}
\usepackage[utf8]{inputenc}
\usepackage{ifpdf}
\usepackage[english]{babel}
\usepackage{caption}
\usepackage{subfig} 
\usepackage{color}
\usepackage{booktabs}
\usepackage{lipsum}

\usepackage{bm}

\newcommand{\field}[1]{\mathbb{#1}} 

\newcommand{\BIGO}[1]{\textrm{O} \left ( {#1} \right )}

\newcommand{\BRAK}[1]{\left [ {#1} \right ]}
\newcommand{\CBRAK}[1]{\left \{ {#1} \right \} }
\newcommand{\ABS}[1]{\left | {#1} \right | }
\newcommand{\PAREN}[1]{\left ( {#1} \right )}

\newcommand{\argmax}[1]{\underset{#1}{\operatorname{arg}\,\operatorname{max}}\;}
\newcommand{\argmin}[1]{\underset{#1}{\operatorname{arg}\,\operatorname{min}}\;}

\newcommand{\NORM}[1]{\left \| {#1} \right \| }

\newcommand{\FOURIER}[1]{\mathcal{F} \BRAK{{#1}} }

\newcommand{\FLOOR}[1]{\left \lfloor {#1} \right \rfloor}
\newcommand{\CEIL}[1]{\left \lceil {#1} \right \rceil}

\newcommand{\MAT}[1]{\bm{{#1}}}
\newcommand{\VEC}[1]{\bm{{#1}}} 

\DeclareMathSymbol{\minus}{\mathbin}{AMSa}{"39}

\DeclareMathOperator{\sinc}{sinc}

\input glyphtounicode
\ninept

\newcounter{numauth}
\newcounter{listcnt}
\newcommand\authcnt[1]{\ifdefined#1 \stepcounter{numauth} \fi}

\newcommand\addauth[1]{
\ifdefined#1 
\stepcounter{listcnt}
\ifnum \value{listcnt}<\value{numauth}
\appto\authorslist{, #1}
\else
\appto\authorslist{~and~#1}
\fi
\fi}
\authcnt{\paperauthorB}
\authcnt{\paperauthorC}
\authcnt{\paperauthorD}
\authcnt{\paperauthorE}
\authcnt{\paperauthorF}
\authcnt{\paperauthorG}
\authcnt{\paperauthorH}
\authcnt{\paperauthorI}
\authcnt{\paperauthorJ}
\def\authorslist{\paperauthorA}
\addauth{\paperauthorB}
\addauth{\paperauthorC}
\addauth{\paperauthorD}
\addauth{\paperauthorE}
\addauth{\paperauthorF}
\addauth{\paperauthorG}
\addauth{\paperauthorH}
\addauth{\paperauthorI}
\addauth{\paperauthorJ}

\usepackage{times}

\newif\ifpdf
\ifx\pdfoutput\relax
\else
   \ifcase\pdfoutput
      \pdffalse
   \else
      \pdftrue
   \fi
\fi

\ifpdf 
  \usepackage[pdftex,
    pdftitle={\papertitle},
    pdfauthor={\authorslist},
    pdfsubject={Proceedings of the 29th International Conference on Digital Audio Effects (DAFx26)},
    colorlinks=false, 
    bookmarksnumbered, 
    pdfstartview=XYZ 
  ]{hyperref}
  \usepackage[pdftex]{graphicx}
\else 
  \usepackage[dvips]{epsfig,graphicx}
  \usepackage[dvips,
    pdftitle={\papertitle},
    pdfauthor={\authorslist},
    pdfsubject={Proceedings of the 29th International Conference on Digital Audio Effects (DAFx26)},
    colorlinks=false, 
    bookmarksnumbered, 
    pdfstartview=XYZ 
  ]{hyperref}
\fi
\usepackage[hypcap=true]{caption}
\title{\papertitle}

\affiliation
{\paperauthorA\, }
{\href{https://nuspaceaudio.com/}{NuSpace Audio} \\ Cambridge, MA, USA\\
{\tt \href{mailto:luoyuancheng@gmail.com}{luoyuancheng@gmail.com}}
}

\begin{document}
\ifpdf 
  \DeclareGraphicsExtensions{.png,.jpg,.pdf}
\else  
  \DeclareGraphicsExtensions{.eps}
\fi


\maketitle

\begin{abstract}

The image-source model (ISM) is a widely adopted method for efficiently simulating acoustic room impulse responses (RIRs) under specular reflection assumptions. Acoustic paths between source and receiver are traced to lattice points computed from successive reflections over bounding planes of the room. Rectangular rooms bound the total number of image-sources to be polynomial in the RIR's duration or distance $k$ equivalent, with degree equal the number of room dimensions $N$. Direct ISM simulations are therefore compute upper-bound by $\BIGO{k^N}$, and consider only cases of $N \leq 3$ for tractability and real-world applications. This work proposes an alternative computational method that lowers the asymptotic compute bound to $\BIGO{N k^2 \log k}$ for integer coordinates and room dimensions via reducing ISM lattice point counting to the classic Gauss circle problem (GCP). We extend the lattice counting model to frequency-dependent and reflection weighted image-sources in higher dimensions, relating solutions between successive dimensions via the convolution operator. Two constructions for realizing RIRs are presented, along with time-frequency controls, error and run-time analysis, and RIR statistics.

\end{abstract}

\section{Introduction}
\label{SEC:INTRO}

The classic image method \cite{allen1979image} for simulating small-room acoustics is widely regarded as a close approximation of the acoustic wave equation, and as an efficient ray-tracer for the class of rectangular shaped rooms \cite{habets2006room}. ISM models specular reflected acoustic ray-paths via vectors between a receiver and a set of image-source coordinates computed from repeated reflections of the source's coordinates over the room boundary's orthogonal planes.
While multiple sequences of reflections can yield the same image-source coordinate, exactly one valid acoustic path exists between receiver and source in the direction and with the path distance of the image-source-receiver vector. 
The number of acoustic paths with distances less than $k$ is the number of unique image-source coordinates bounded by a $k$-radius ball, and therefore is proportional to the volume $\BIGO{k^N}$ of the bounding sphere in $N$ number of dimensions.
This fact is not shared by source-receivers in non-rectangular enclosures where image-sources are seldom coincident and represent valid acoustic paths  \cite{borish1984extension},  \cite{rindel2000use}. 
Thus, ISM's efficacy depends on the regularity of rectangular rooms, forming a so-called lattice in $\field{R}^N$ that bounds each image-source.

This work proposes an alternative computational method for ISM, which exploits the lattice topology when summing image-source contributions for RIR generation, and drastically reduces computational costs for $N \geq 3$ dimensions in specific conditions.
We assume image-sources contribute only to the delay, distance attenuation, and wall-reflection components of the acoustic response. Incidence angle and distances from receiver to specific image-sources are lost, which forecloses direct extension of some works in literature:
Acoustic directivity of sources and receivers remain omni-directional unlike generalizations to microphone and loudspeaker directivity in \cite{xu2024simulating}, \cite{luo2021fast}, \cite{samarasinghe2018spherical}, and binaural RIRs \cite{heinz1993binaural}.
Wall-reflection coefficients may be complex for modeling its acoustic impedances, but are assumed to be independent of incident angles unlike \cite{aretz2014application}.
Furthermore, we confine source and receiver coordinates, and room dimension to the integers in $\field{Z}^N$, which no longer suites continuous moving sources and receivers \cite{ali2025source, schissler2011gsound}.

Our motivations are twofold: First, we observe in the acoustic wave equation's solution that the number of eigenfrequencies increases with the number of dimensions of a rectangular room \cite{kuttruff2016room_normalmodes}. This induces higher modal densities in ISM's reverberation tail, which reduces unwanted coloration.
However, conventional ISM reverberation tails are often omitted due to prohibitive cubic $\BIGO{k^3}$ run-time costs of computing high order reflections. Instead, hybrid RIR models  \cite{lehmann2009diffuse}, \cite{thomas2017wayverb} use efficient alternatives such as feedback delay networks \cite{gerzon1971synthetic}, \cite{jot1991digital}, \cite{schlecht2016lossless} to produce dense but colored reverberation tails when not optimized.
High-dimensional meshes for FDTD room simulations \cite{kelloniemi2005artificial} can increase modal density to reduce coloration but have high computational costs proportional to  the product of cells per dimension.
Our model's computational complexity is $\BIGO{N k^2 \log k}$, which improves upon the $\BIGO{k^N}$ costs of direct $N$-dimensional ISM in \cite{mcgovern2011image}, and able to produce both higher-order and denser reflections for larger $k$ and $N$.
Second, late-reverberation tails in large real-world spaces take on statistical properties of Gaussian densities, with exponentially increasing number of reflections \cite{jot1997analysis}, \cite{georganti2008analysis}.
However, ISM's number of reflections has polynomial growth, and the RIRs generated from \cite{scheibler2018pyroomacoustics}, \cite{habets2006room} have reverberation times that differ from well-established predictions \cite{lehmann2008prediction} such as those made by Sabine \cite{sabine2015collected} and Eyring \cite{eyring1930reverberation}. 
We therefore report on statistical properties of ISM's RIRs such as echo density \cite{abel2006simple}, energy decay curve (EDC), P50 energy percentile, and estimated RT60 with its coefficient of determination (Rsq). Our implementation is available online\footnote{\url{https://github.com/yluo1/GCP-ISM}}, and paper is organized as follows:

Section \ref{SEC:GCP} relates ISM to the so-called Gauss circle problem (GCP), which we efficiently generalize for $N$ dimensions via recurrence relations and express via the convolution operation.
Section \ref{SEC:ISM_GCP} extends GCP's volume function to support lattice point translation, scaling, and weighting, which models image-source coordinates and exponentiated wall-reflection coefficients.
Section \ref{SEC:FOR_INV_ISM_GCP} presents two methods for recovering the RIR via differentiating GCP-ISM's volume function.
Section \ref{SEC:TIME_FREQ} combines GCP-ISM simulations computed over complex reflection coefficients to recover a RIR.
We perform a case-study of echo density profiling, along with error and run-time performance analysis in section \ref{SEC:EXP}.
Section  \ref{SEC:CONCLUSION} summarizes our findings and concludes the work.

\section{Image-Source Gauss Circle Problem} 
\label{SEC:GCP}

\begin{figure}[tb]
\centerline{\includegraphics[width=0.8\columnwidth]{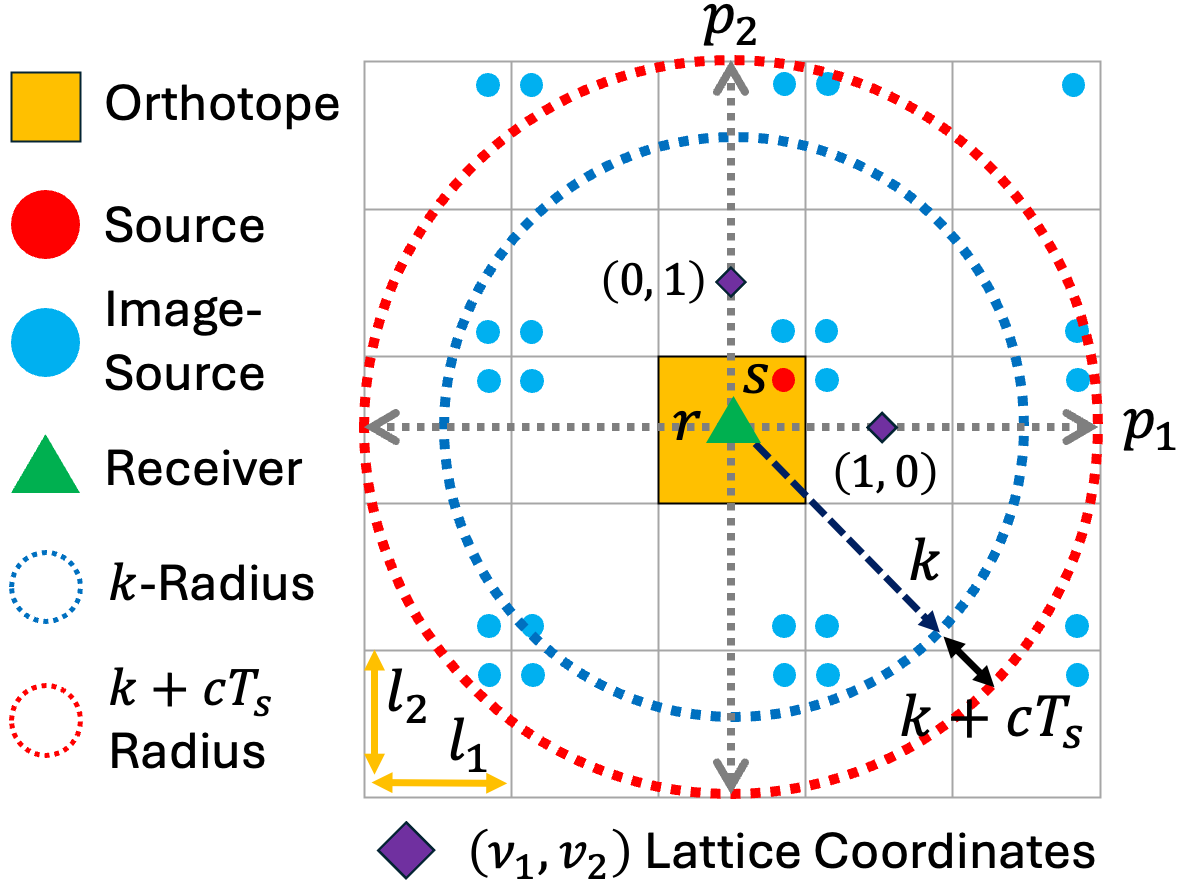}}
\caption{\label{FIG:ISM}{Reflecting the source across the orthotope planes generates the image-sources. Image-sources between $k$ and $k+c T_s$ contribute to the RIR at time $k/c$ with delay and attenuation proportionate to their distances $ \NORM{\VEC{p}(\VEC{\nu}) - \VEC{r} }_2$ from the receiver at $\VEC{r}$.}}
\end{figure}

The reflected room boundaries in ISM are defined by a lattice of cuboids, and generalized to be $N$-dimensional orthotopes over a set of lattice coordinates. Let a lattice coordinate be given by
\begin{equation} \label{EQ:LATTICE}
\displaystyle
\begin{split}
\VEC{\nu} =  \BRAK{\nu_1, \hdots, \nu_N}^T, \quad \nu_n \in \field{Z},
\end{split}
\end{equation}
where the lattice coordinate origin $\VEC{\nu} = \VEC{0}$ indexes the orthotope of size $\VEC{\ell} = \BRAK{\ell_1, \hdots, \ell_N}^T  \in \field{R}^N$ containing an omni-directional sound source at $\VEC{s} = \BRAK{s_1, \hdots, s_N}^T   \in \field{R}^N$ and receiver at $\VEC{r} = \BRAK{r_1, \hdots, r_N}^T  \in \field{R}^N$ in meters. Specular acoustic paths between source, boundaries, and receivers  are modeled by successive reflections or mirroring of the source across the orthotope's planes. For brevity, we can express the image-sources' coordinates as a function of the lattice coordinates  $\VEC{p}(\VEC{\nu}) = \BRAK{p_1(\nu_1), \hdots, p_N(\nu_N) }^T$ as shown in Fig. \ref{FIG:ISM}, where the $n^{th}$ coordinate is given by
\begin{equation} \label{EQ:IMAGE_COORDS}
\displaystyle
\begin{split}
p_n(\nu_n) = \ell_n \nu_n + (-1)^{\nu_n} s_n,
\end{split}
\end{equation}
and where the acoustic path-length equates to the image-source to receiver distance $d(\VEC{\nu}) = \NORM{\VEC{p}(\VEC{\nu}) - \VEC{r} }_2$. Therefore, the set of all lattice coordinates with image-sources within time $T$ (seconds) is given by $\Xi = \CBRAK{\VEC{\nu}  \in \field{Z}^N  \, | \, \NORM{d(\VEC{\nu})}_2 \leq c T  }$, where $c$ is the speed of sound in meters / second.

\setlength{\textfloatsep}{8pt}  

The image-source model's acoustic room transfer function (RTF) $H(\omega)$ at angular frequency $\omega$ between source and receiver integrates over all image-sources and is given by 
\begin{equation} \label{EQ:RTF} 
\displaystyle
\begin{split}
H(\omega) &  = \sum_{\VEC{\nu} \in \Xi }  A(\VEC{\nu}) R(\omega, \VEC{\nu})  e^{\frac{\minus j \omega d(\VEC{\nu})}{c T_s} }, \quad
A(\VEC{\nu})  = \frac{1}{d^{ \frac{N \minus 1}{2}}(\VEC{\nu}) }, \\ 
R(\omega, \VEC{\nu}) & = \prod_{n=1}^N \Gamma_{+n}^{\ABS{\FLOOR{\frac{\nu_n + 1}{2}}}}(\omega) \, \Gamma_{\minus n}^{\ABS{\FLOOR{\frac{1 \minus \nu_n}{2}}}}(\omega),
\end{split}
\raisetag{5ex}
\end{equation}
where $T_s$ is the sampling period (seconds), and $\FLOOR{x}$, $\CEIL{x}$ are the floor and ceiling functions respectively. The distance attenuation term $A(\VEC{\nu})$ normalizes for total energy scattered over the surface of a $N$-dimensional sphere. The wall-material transfer functions $\Gamma_{+n}(\omega)$, $\Gamma_{\minus n}(\omega)$ model reflections between positive and negative facing walls in the $n^{th}$ dimension respectively; $R(\omega, \VEC{\nu})$ models the contribution of all wall-reflections for the image-source at $p(\VEC{\nu})$. Lastly, the complex exponential in the frequency domain accounts for the acoustic delay. Thus, the direct cost of computing Eq. \eqref{EQ:RTF} containing up to the $K^{th}$ order reflection is polynomial $\BIGO{K^N}$, which is prohibitive for computing both the RTF's reverberation tail, and generalizing into higher-dimensional rooms ($N > 3$). Fortunately, we can show the integration to be separable across successive dimensions as follows:

Let us consider integrating image-source contributions over radial slices of $\field{R}^N$ consisting of contributions from the lower dimensional slice in $\field{R}^{N \minus 1}$. To motivate and then illustrate this approach, we introduce the Gauss circle problem, which counts the number of integer lattice coordinates within a $k$-radius ball under the Euclidean norm $\NORM{\VEC{\nu}}_{2} = \sqrt{ \sum_{n=1}^N \nu^2_n}$ as shown in Fig. \ref{FIG:GCP}. Source and receiver coordinates are coincident to the origin within a unit sized orthotope; image-source coordinates are coincident to integer lattice coordinates. Thus, discretizing GCP's radius $k$ over integer multiples of the sampling distance $c T_s$ and taking the difference between consecutive samples yields the number of image-sources in the sampling interval. Taking weighted finite differences of GCP w.r.t. $k$ therefore approximates the RIR of Eq. \eqref{EQ:RTF}. We formulate the discretized GCP volume function as follows: 

\begin{figure}[tb]
\centerline{\includegraphics[width=0.65\columnwidth]{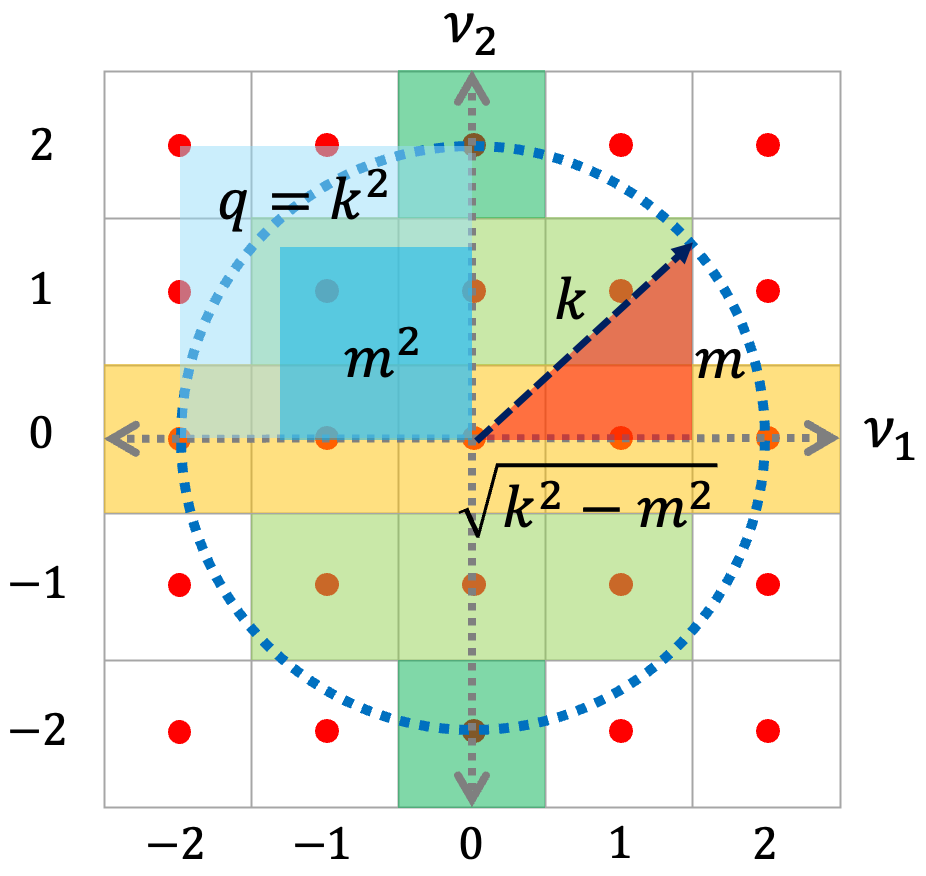}}
\caption{\label{FIG:GCP}{The number of lattice points in row $\nu_2 = m$ within the $k$-radius circle equals that of the preceding dimension $\nu_1$ of smaller radius  $\sqrt{k^2 - m^2}$. Efficient memoization indexes the differences of squares or area $q$ within the square-root term.}}
\end{figure}

Let $C(k, N) = \ABS{ \CBRAK{\VEC{\nu} \in \field{Z}^N \, \left | \,  \NORM{\VEC{\nu}}_2 \leq k \right .  } } $ be the cardinality of the set of lattice coordinates, upper bounded by $k$ Euclidean norm. Squaring and subtracting the $N^{th}$ coordinate $\nu_N$ from both sides of the inequality yields the equivalent set and size given by
\begin{equation} \label{EQ:GCP}
\displaystyle
\begin{split}
C(k, N) & = \ABS{ \CBRAK{\VEC{\nu} \in \field{Z}^N \, \left | \,  \sum_{n=1}^N \nu_n^2  \leq k^2  \right . } }  \\
& = \sum_{\nu_N = \minus \infty}^{ \infty}   \ABS{ \CBRAK{\VEC{\nu} \in \field{Z}^{N \minus 1} \, \left | \,  \sum_{n=1}^{N \minus 1} \nu_n^2  \leq k^2 - \nu_N^2  \right . } } \\
& =  \sum_{\nu_N = \minus \infty}^{ \infty} C \PAREN{ \sqrt{k^2 - \nu_N^2}, \, N-1}.
\end{split}
\end{equation}
The set size therefore has a recurrence relation that integrates over the cardinality of the set of lattice points $\VEC{\nu} \in \field{Z}^{N \minus 1}$ in the lower preceding dimensions as shown in Fig. \ref{FIG:GCP}. Computing GCP's base-case ($N=1$) and higher dimension cases ($N > 1$) gives
\begin{equation} \label{EQ:GCP_COUNT}
\displaystyle
\begin{split}
C(k, N)  & = \left \{ \begin{array}{cc}
 1 + 2 \FLOOR{k}, & N = 1 \\
 \sum \limits_{m = \minus \FLOOR{k} }^{\FLOOR{k}}  C \PAREN{\sqrt{k^2 - m^2}, \, N-1}, &  N > 1\end{array} \right . 
\end{split} \raisetag{7ex}
\end{equation}
by which the summation bounds are tightened. 

\begin{figure}[tb]
\centerline{\includegraphics[width=0.99\columnwidth]{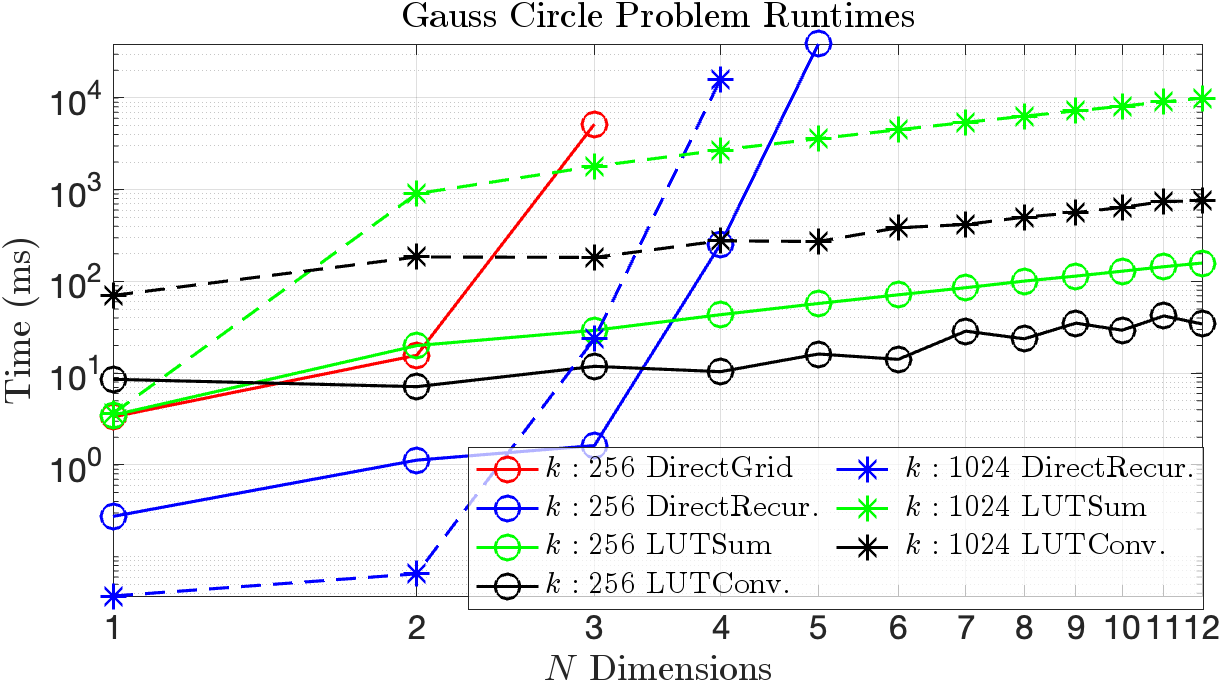}}
\caption{\label{FIG:GCP_RUNTIMES}{GCP run-times on Mac M1 Matlab approach asymptotic growth for small $k$ and $N$. Run-times are prohibitive for direct grid Eq. \eqref{EQ:GCP} at $N=3$, and direct recurrence Eq. \eqref{EQ:GCP_COUNT} at $N=4$. LUT summation Eq. \eqref{EQ:GCP_COUNT_MEMO} and convolution Eq. \eqref{EQ:GCP_COUNT_MEMO_CONV_FFT} are much faster, where the latter completes in under $1$ second for all dimensions $N \leq 12$.  }}
\end{figure}

Observe that for integer distances $k$, we can memoize sub-solutions of Eq. \eqref{EQ:GCP_COUNT} by indexing the integer area $q \in \field{Z}_{\geq 0}$ of the square-root term, and evaluate $C \PAREN{ \sqrt{q}, N} = \hat{C}(q, N)$ over a look-up-table (LUT) populated by the following area function:
\begin{equation} \label{EQ:GCP_COUNT_MEMO}
\displaystyle
\begin{split}
\hat{C}(q, N) & = \left \{
\begin{array}{cc}
1 + 2 \FLOOR{\sqrt{q}}, 					 & N = 1 \\ [2pt]
 \sum \limits_{m = \minus \FLOOR{\sqrt{q}} }^{\FLOOR{\sqrt{q}}}  \hat{C}\PAREN{{q - m^2}, N - 1}, &  N > 1
\end{array}
\right.
\end{split}
\end{equation}
where $q = k^2$, and $\hat{C}(q, N)$ requires $\BIGO{N k^2}$ memory and $\BIGO{N k^3}$ compute costs under direct evaluation. We can reduce the computation costs of the summation in Eq. \eqref{EQ:GCP_COUNT_MEMO} w.r.t. $k$ for $N > 1$ by expressing the recurrence relation in terms of the equivalent geometric convolution operation $\hat{C}_{N} = f *\hat{C}_{N \minus 1}$ given by
\begin{equation} \label{EQ:GCP_COUNT_MEMO_CONV}
\displaystyle
\begin{split}
\hat{C}(q, N) & = 
 \sum_{\tau = 0 }^{k^2}  f(\tau) \,  \hat{C}\PAREN{q - \tau, N - 1}, \\
 f(\tau) & =  \left \{  \begin{array}{cl} 1, & \tau = 0 \\ 2, & \tau \in \CBRAK{n^2 \, | \, n \in \field{N} }  \\ 0, & \textrm{Otherwise} \end{array} \right . ,
\end{split}
\end{equation}
where $f(\tau)$ is a sparse convolution kernel with non-zero values at indices that are perfect squares. Using the Fast Fourier transform (FFT) $\mathcal{F}$ with zero-padding to $N k^2$ elements, and element-wise product $\odot$ operators, the recurrence relation simplifies to multiplication in the frequency domain given as follows:
\begin{equation} \label{EQ:GCP_COUNT_MEMO_CONV_FFT}
\displaystyle
\begin{split}
\FOURIER{\hat{C}_N} & = \FOURIER{f} \odot  \FOURIER{\hat{C}_{N \minus 1}} 
 = \mathcal{F}^{N \minus 1} \BRAK{f} \odot  \FOURIER{\hat{C}_{1}},
\end{split}
\end{equation}
whereby $f = \BRAK{f(0), \, \hdots, \, f(k^2)}$ in the frequency domain undergoes element-wise exponentiation, and $\hat{C}_N$ is recovered via inverse FFT. Thus, the computation costs are further reduced to $\BIGO{N k^2 \log k}$. Sample run-time performances across the methods in Fig. \ref{FIG:GCP_RUNTIMES} show that the asymptotic complexity dominates the costs due to the relative simplicity of the formulations. Next, we introduce our GCP-ISM extension, which follows an analogous derivation of Eq. \eqref{EQ:GCP} and modifications to Eqs. \eqref{EQ:GCP_COUNT_MEMO},  \eqref{EQ:GCP_COUNT_MEMO_CONV}, \eqref{EQ:GCP_COUNT_MEMO_CONV_FFT}.

\subsection{GCP with Image-Source Extensions} \label{SEC:ISM_GCP}
Let $S(k, N) = \ABS{ \CBRAK{\VEC{\nu} \in \field{Z}^N \, \left | \,  \NORM{\VEC{p}(\VEC{\nu}  ) - \VEC{r}  }_2 \leq k \right .  } } $  be the cardinality of the set of source and image-sources with coordinates $\VEC{p}(\VEC{\nu})$ within $k$ Euclidean distance units from the receiver coordinate $\VEC{r}$. Expanding and rearranging the inequality terms as in Eq. \eqref{EQ:GCP}, and substituting $p_n(\nu_n)$ in Eq. \eqref{EQ:IMAGE_COORDS} yields the recurrence relation between set sizes across lower preceding dimensions given by
\begin{equation} \label{EQ:IS_GCP} 
\displaystyle
\begin{split}
S(k, N) & = \ABS{ \CBRAK{\VEC{\nu} \in \field{Z}^N \, \left | \,  \sum_{n=1}^N \PAREN{p_n(\nu_n) - r_n}^2  \leq k^2  \right . } }  \\
& =  \sum_{\nu_N = \minus\infty}^{ \infty} S \PAREN{ \sqrt{k^2 - \PAREN{p_N(\nu_N) - r_N} ^2}, \, N-1}.
\end{split}
\raisetag{12ex}
\end{equation}
Unlike GCP, the source and receiver coordinates are not coincident to lattice coordinates when translated away from the origin. Moreover, tight upper and lower bounds of the summation in Eq. \eqref{EQ:IS_GCP} may be asymmetric as successive image-source coordinates reflect over planes as seen in Fig. \ref{FIG:ISM}.
Applying the recurrence-relation of Eq. \eqref{EQ:GCP_COUNT} and GCP area function in Eq. \eqref{EQ:GCP_COUNT_MEMO} yields the GCP-ISM area function $S\PAREN{ \sqrt{q}, N} = \hat{S}(q, N)$  for $q \in \field{R}_{\geq 0}$ given by
\begingroup
\setlength{\arraycolsep}{3pt}
\begin{equation} \label{EQ:IS_GCP_COUNT_MEMO}
\displaystyle
\begin{split}
\hat{S}(q, N) & = \left \{
\begin{array}{cc}
1 + b(q) - a(q), 	 &  \begin{array}{c} N = 1, \\ b(q) \geq a(q)  \end{array} \\ [8pt] 
\sum \limits_{m = a(q)}^{b(q) }  \hat{S}\PAREN{{q - u^2(m, N)}, \, N-1}, &   \begin{array}{c} N >  1, \\ b(q) \geq a(q)  \end{array} \\ [10pt] 
0, 		&  \textrm{Otherwise} \\
\end{array}
\right. \\
u(m, N) & =  \ell_N m + (-1)^m s_N - r_N, \quad \textrm{via Eq. \eqref{EQ:IMAGE_COORDS}}
\end{split}
\raisetag{2.75ex}
\end{equation}
\endgroup
where we sum over lattice indices $m$ between upper and lower bounds $b(q)$, $a(q)$ respectively that cover the signed distance $u(m, N)$ between the image-source and receiver along the $N^{th}$ dimension. 
The bounds follow from the constraint $\ABS{u(m, N)} \leq \sqrt{q}$ that contain the source-receiver distance $u(0, N)$, and satisfies
\begin{equation} \label{EQ:IS_GCP_COUNT_MEMO_BOUNDS}
\displaystyle
\begin{split}
-\sqrt{q} & \leq u\PAREN{a(q), N} \leq  u(m, N) \leq u\PAREN{b(q), N}  \leq \sqrt{q}, \\
b(q) &  = \bar{g} \PAREN{\frac{r_N + \sqrt{q}}{\ell_N}, \, \frac{s_N}{ \ell_N } }, \quad \qquad \textrm{Upper bound} \\
a(q) & = \underline{g} \PAREN{\frac{r_N - \sqrt{q} }{\ell_N}, \, \frac{s_N}{ \ell_N } },  \qquad \quad \textrm{Lower bound} 
\end{split}
\end{equation}
where the upper and lower extrema functions $\bar{g}(x,y)$ and $\underline{g}(x,y)$ are defined in Appendix Eqs. \eqref{EQ:APP:MAX_INT}, \eqref{EQ:APP:MIN_INT} respectively. Weighting the image-source summations in Eq. \eqref{EQ:IS_GCP_COUNT_MEMO} by the reflection coefficients in Eq. \eqref{EQ:RTF} gives the following GCP-ISM volume function:
\begingroup
\setlength{\arraycolsep}{3pt}
\begin{equation} \label{EQ:IS_GCP_MEMO_WT}
\displaystyle
\begin{split}
\ddot{S}(q, N) & =  \sum \limits_{m = a(q)}^{b(q) } 
 \left \{ \begin{array}{cc} 
1, &  \begin{array}{c} N = 1, \\ b(q) \geq a(q)  \end{array} \\ [10pt] 
 \ddot{S}\PAREN{{q - u^2(m, N) }, \, N-1}, &  \begin{array}{c} N >  1, \\ b(q) \geq a(q)  \end{array} \\ [10pt] 
 0, 		&  \textrm{Otherwise} \\
 \end{array}
\right .  \\
& \times W(m, N), 
\quad W(m, N) = \Gamma_{+N}^{\ABS{\FLOOR{\frac{m + 1}{2}}}} \, \Gamma_{\minus N}^{\ABS{\FLOOR{\frac{1 \minus m}{2}}}}, 
\end{split}
\raisetag{3ex}
\end{equation}
\endgroup
where $\Gamma_{+ N}, \, \Gamma_{\minus N}$ are complex reflection coefficients. We now show that the one-dimensional case is analytic, and the higher-dimensional cases are convolution of lower-dimension solutions.

The base case ($N = 1$) of Eq. \eqref{EQ:IS_GCP_MEMO_WT} can be computed via accumulation to yield the sum of geometric series of the parity and signed components given by
\begin{equation} \label{EQ:IS_GCP_MEMO_WT_REFL_BASE}
\displaystyle
\begin{split}
\ddot{S}(q, 1) & =  \frac{1 - \Gamma_{N}^{\CEIL{\ABS{\frac{b(q) + 1}{2}}}  }  + \PAREN{1 - \Gamma_{N}^{\CEIL{\ABS{\frac{b(q )}{2}}}  }  }  \Gamma_{+N} }{1 - \Gamma_{N}} \\
& +  \frac{1 - \Gamma_{N}^{\CEIL{\ABS{\frac{a(q) \minus 1}{2}}}  }  + \PAREN{1 - \Gamma_{N}^{\CEIL{\ABS{\frac{a(q)}{2}}}  }  }  \Gamma_{\minus N} }{1 - \Gamma_{N}}
 - 1, \\
\end{split}
\end{equation}
where $\Gamma_{N} = \Gamma_{+ N} \Gamma_{\minus N}$, and $\Gamma_{N} \neq 1$. The high-dimensional cases ($N>1$) can be efficiently memoized for integer input parameters; distances $k, \, q \in \field{Z}_{\geq 0}$, orthotope dimensions $\VEC{\ell} \in \field{Z}_{\geq 0}^{N}$, source and receiver coordinates $\VEC{s}, \, \VEC{r} \in \field{Z}^{N} $ ensures that $q - u^2(m, N) \in \field{Z}_{\geq 0}$ is a non-negative index. This allows Eq. \eqref{EQ:IS_GCP_MEMO_WT} to be expressed in the geometric convolution form of Eq. \eqref{EQ:GCP_COUNT_MEMO_CONV} given by
\begin{equation} \label{EQ:IS_GCP_MEMO_WT_REFL_CONV}
\displaystyle
\begin{split}
\ddot{S}(q, N) & = 
 \sum_{\tau = 0 }^{k^2}  \PAREN{ \bar{f}(\tau, N) +  \underline{f}(\tau, N)  }  \ddot{S}\PAREN{q - \tau, N - 1}, \\
\bar{f}(\tau, N) & =   \left \{  \begin{array}{cl}
  W (m, N), & \tau = u^2(m, N) , \, m \geq 0  \\ 
  0, & \textrm{Otherwise} \end{array} \right . , \\
 \underline{f}(\tau, N) & =   \left \{  \begin{array}{cl}
  W (m, N) , & \tau   = u^2(m, N), \,  m < 0  \\ 
  0, & \textrm{Otherwise} \end{array} \right . , \\
\end{split}
\end{equation}
where the convolutions can be efficiently computed via FFTs and inverse FFT following Eq. \eqref{EQ:GCP_COUNT_MEMO_CONV_FFT} with zero-padding to $N k^2$ elements:
\begin{equation} \label{EQ:IS_GCP_COUNT_MEMO_CONV_FFT}
\displaystyle
\begin{split}
\FOURIER{\hat{S}_N} & = \FOURIER{f_N} \odot  \FOURIER{\ddot{S}_{N \minus 1}} , \\
\end{split}
\end{equation}
where $f_N  = \BRAK{\bar{f}(0, N) + \underline{f}(0, N), \, \hdots, \, \bar{f}(k^2, N) + \underline{f}(k^2, N) }$. Thus, computing the LUT $\ddot{S}(q, N)$ for integer inputs only requires $\BIGO{N k^2 \log k}$, which is linear w.r.t. the number of dimensions $N$. For comparison, directly computing $\ddot{S}(q, N)$ without memoization requires $\BIGO{k^{N \minus 1}}$. Computing direct ISM from Eq. \eqref{EQ:RTF} up to max-order reflections $M = k / \min \CBRAK{\ell_1, \, \hdots, \, l_N}$ that fits the axial reflections along the shortest orthotope dimension within $k$ meters requires $\BIGO{k^N}$. Next, we give two constructions of GCP-ISM RIRs from the LUTs via finite-difference operators.

\subsection{GCP-ISM Forward and Inverse RIR Constructions} \label{SEC:FOR_INV_ISM_GCP}

For the forward construction of RIR $h(i)$, we can compute the $i^{th}$ sample at time $T_s i$ from the finite-difference of the GCP-ISM volume function in Eq. \eqref{EQ:IS_GCP_MEMO_WT}  w.r.t. sample distances $k_i$ given by
\begin{equation} \label{EQ:SAMPLING_FORWARD}
\displaystyle
\begin{split}
h(i) & =  \frac{\ddot{S} \PAREN{k_{i+1}^2, \,  N} - \ddot{S} \PAREN{k_{i}^2, \, N} }{k_{i}^{\frac{N \minus 1}{2}}} , \quad
k_i  = c T_s i,
\end{split}
\end{equation}
where $k_i$ is proportional to the sample time and sound speed, and the denominator component normalizes for spherical scattering energy. Most sample distances $k_i$ are not integers, and the non-uniform distance resolution of values stored in $\ddot{S}(q, N)$ for integers $q$ is quadratic w.r.t. $k$; fewer entrants are represented for smaller $k$. We therefore expect larger errors when rounding up the squared distance and evaluating the volume function $\ddot{S} \PAREN{\CEIL{k_i^2}, \,  N}$ for smaller $k_i$. Fortunately, we observe that GCP-ISM function $S(k, N)$  in Eq. \eqref{EQ:IS_GCP}  is invariant to coordinate-scaling; up-scaling the inputs $\lambda k$, $\lambda \VEC{\ell}$, $\lambda \VEC{s}$, $\lambda \VEC{r}$ by $\lambda \in \field{Z}_{>0}$, yields the equivalent $\lambda$-scaled cardinality functions  $S(k, N) = S^{\CBRAK{\lambda}}(\lambda k, N)$ given by
\begin{equation} \label{EQ:GCP_ISM_SCALE_EQUIV}
\displaystyle
\begin{split}
S^{\CBRAK{\lambda}}(\lambda k, N) = \ABS{ \CBRAK{\VEC{\nu} \in \field{Z}^N \, \left | \,  \NORM{ \lambda \PAREN{ \VEC{p}(\VEC{\nu}  ) - \VEC{r} } }_2 \leq \lambda  k \right .  } },
\end{split}
\end{equation}
and equates their volume functions $\ddot{S}(q, N) = \ddot{S}^{\CBRAK{\lambda}}(\lambda^2 q, N)$. Thus, decreasing the unit-distance by a factor of $\lambda$ increases the distance resolution of GCP-ISM's volume LUT by a factor of $\lambda^2$.

\begin{figure}[tb]
\centerline{\includegraphics[width=0.95\columnwidth]{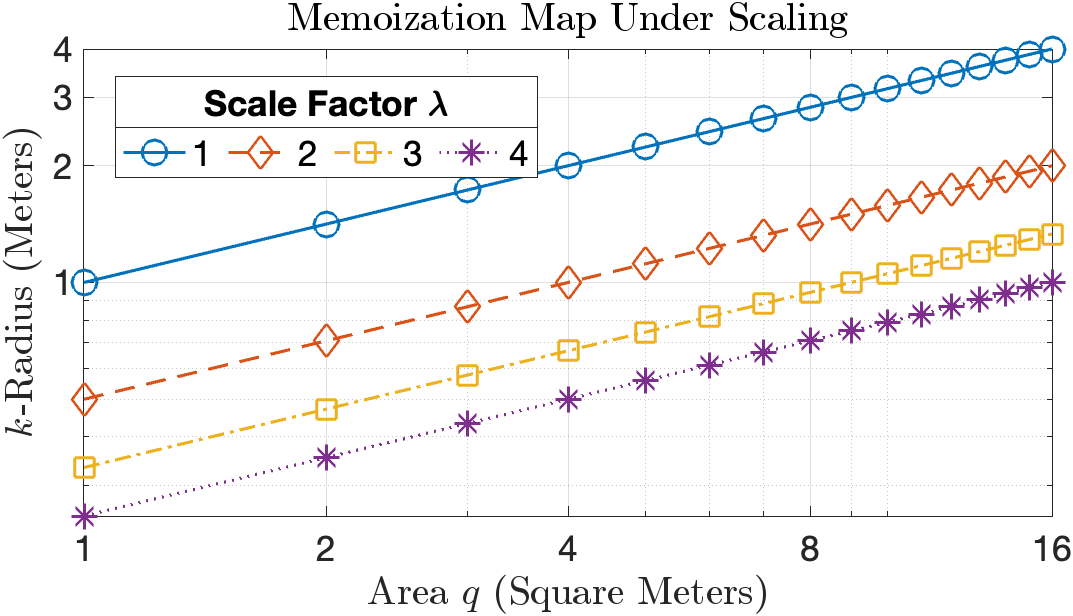}}
\caption{\label{FIG:MEMO_MAP}{The non-scaled ($\lambda = 1$) LUT $\ddot{S}(q, N)$ cannot represent distances below $1$. Scaled $\lambda = \CBRAK{2, \, 3, \, 4}$ LUTs represent $2^\lambda$ distances for $k \leq 1$.  LUT values at $\ddot{S}^{\CBRAK{\lambda}}(\lambda^2 q, N)$ are equivalent. }}
\end{figure}

\begin{figure*}[tb]
\centering 
    \subfloat[Forward construction from Eq. \eqref{EQ:SAMPLING_FORWARD_LUT} with LUT threshold $K_E = 0$ \label{FIG:FORWARD_GCP}]{%
        \includegraphics[width=0.495\textwidth]{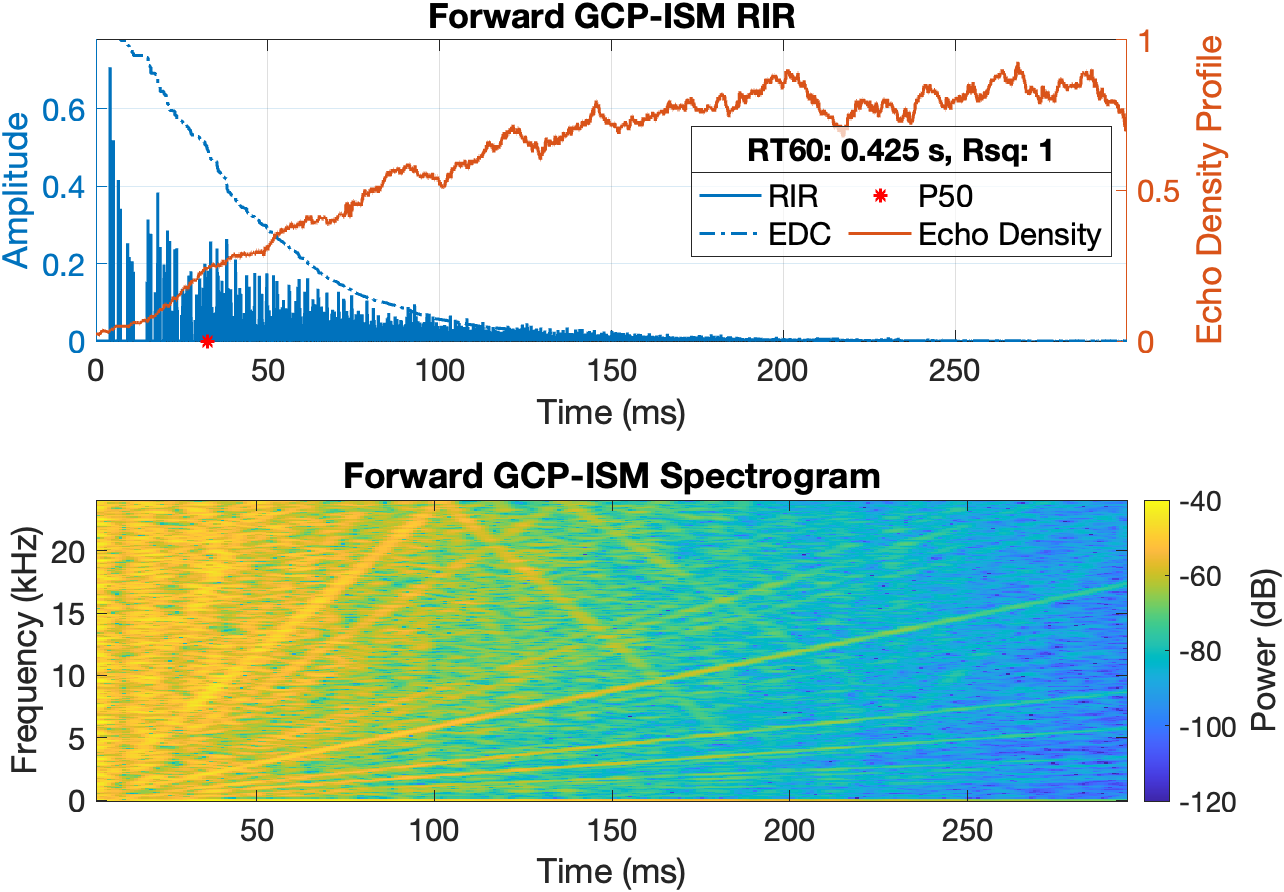}%
    }
    \subfloat[Inverse construction from Eq. \eqref{EQ:SAMPLING_INV_LUT} with Lanczos kernel size $\alpha = 10$  \label{FIG:INVERSE_GCP}]{%
        \includegraphics[width=0.495\textwidth]{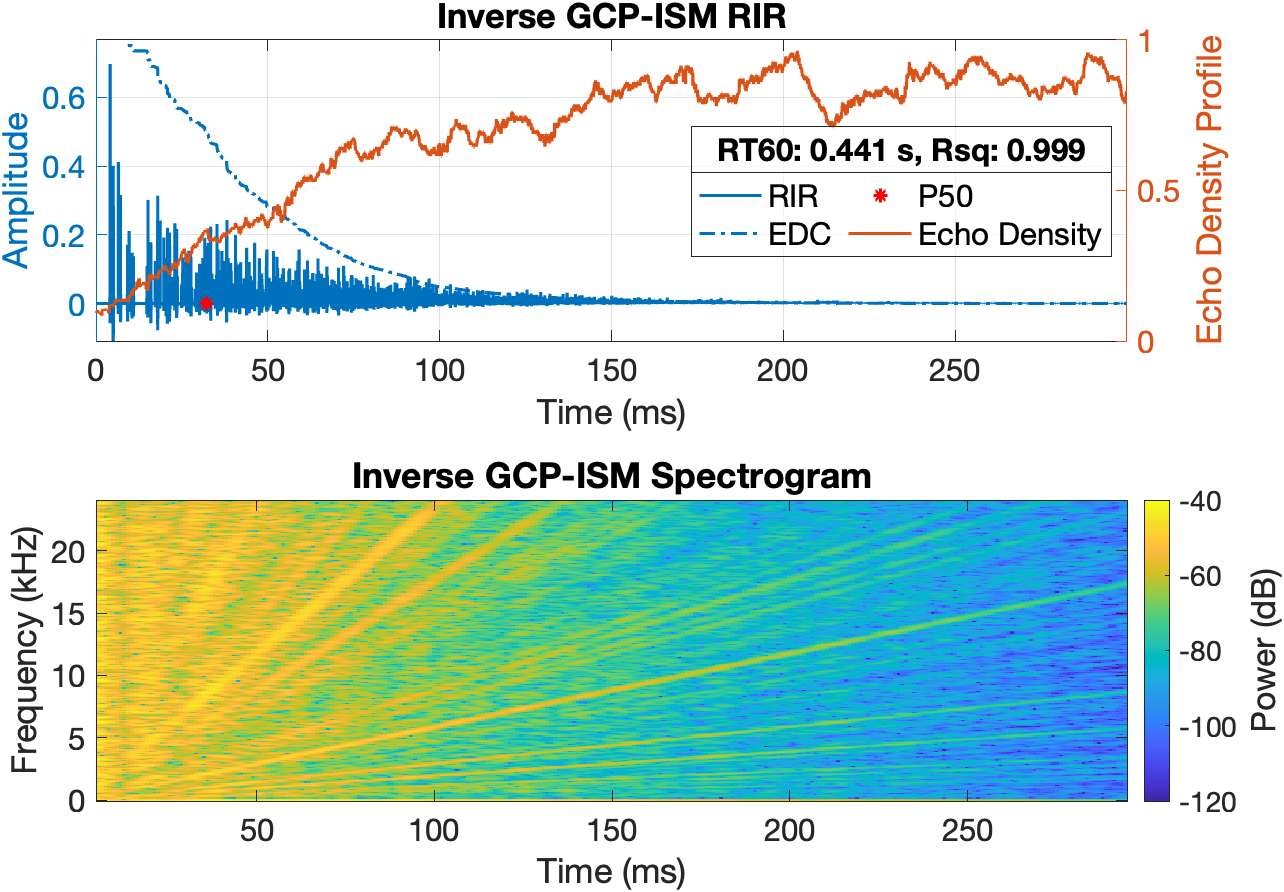}%
    }
      \caption{\label{FIG:FORWARD_INVERSE_GCP} The following parameters generate RIRs for forward and inverse GCP-ISM constructions: $\VEC{s} = \BRAK{1, 0, 1}$, $\VEC{r} = \BRAK{2, 1, 1}$, $\VEC{\ell} = \BRAK{5, 4, 3}$, $\VEC{\Gamma_{+}} = \BRAK{0.93, 0.8,  0.9}$, $\VEC{\Gamma_{\minus}} = \BRAK{0.72, 0.78, 0.8}$, $\lambda = 1$, and $T = 0.3$. The echo density profile $\eta(t)$ in \cite{abel2006simple} tracks the fraction of samples that lie outside the standard deviation of a rectangle sliding window ($512$ samples at $48$ kHz sampling rate). $\eta(t) = 1$ is Gaussian.}    
\end{figure*}

The memoization density map in Fig. \ref{FIG:MEMO_MAP} shows the sample distances $k$ represented by the integers $q$ for varying $\lambda$-scaled LUTs. The number of distances $k$ represented between $k_i \leq k \leq k_j$  by $\lambda$-scaled LUT follows $\lambda^2 \PAREN{k_j^2 - k_i^2 }$. Therefore, we obtain higher resolution at large distances $k$ for $\lambda > 1$ at the cost of increasing the LUT size by a factor of $\lambda^2$. We also note that direct evaluation of $\ddot{S}(k_i^2, N)$ for small distances $k_i$ is inexpensive below a nominal threshold $k_{E}$. It's more accurate to construct the RIR from LUTs at sample distances exceeding $k_{E}$ as follows:
\begin{equation} \label{EQ:SAMPLING_FORWARD_LUT}
\displaystyle
\begin{split}
\ddot{h}(i, \, N) & =   \frac{1}{k_{i}^{\frac{N \minus 1}{2}}}  \left \{ \begin{array}{cc}
  \ddot{S}^{\CBRAK{\lambda}}(q_{i+1}, N)  -   \ddot{S}^{\CBRAK{\lambda}}(q_{i}, N)  , & k_i > k_{E} \\[8pt]
\ddot{S}(k_{i+1}^2, N) - \ddot{S}(k_{i}^2, N), & k_i \leq k_{E} \\
\end{array} \right .
\end{split}
\end{equation}
where we sample the LUT at $\lambda$-scaled squared distances $q_i =  \lambda^2 \CEIL{k_i^2}$ w.r.t. consecutive $k_i$. This construction is comparable to realizing the RTF of Eq. \eqref{EQ:RTF} for real-valued reflection coefficients $\Gamma$, and  without fractional delay supports; complex reflection coefficients would induce variable group-delay, and sample delays are rounded up to the nearest integer. The resulting RIR in Fig. \ref{FIG:FORWARD_GCP} is strictly non-negative, and contains aliasing in the spectrogram. Other characteristics typical of ISM RIRs are reproduced, such as flutter reflections, sweep streaks in the spectrogram, and a rising echo density profile. Next, we support fractional-delays by inverting the sampling direction to map consecutive $q$ to fractional $k$.

For the inverse construction of RIR $h(i)$ up to max distance $\bar{k}$, let us integrate a sinc kernel centered on $i$ over the derivative of the GCP-ISM volume function in Eq. \eqref{EQ:IS_GCP_MEMO_WT} w.r.t. area $q$ as follows:
\begin{equation} \label{EQ:SAMPLING_INV}
\displaystyle
\begin{split}
h(i) & =   \frac{1}{k_{i}^{\frac{N \minus 1}{2}}}  \int_{q = 0}^{\CEIL{\bar{k}^2}} \frac{d \, \ddot{S} \PAREN{q, \,  N} }{dq}  \sinc \PAREN{i - \frac{\sqrt{q}}{c T_s}} \, dq.
\end{split}
\end{equation}
We can approximate the integral by taking the finite-difference of the memoized and $\lambda$-scaled volume function in Eq. \eqref{EQ:GCP_ISM_SCALE_EQUIV} w.r.t. $q$:
\begin{equation} \label{EQ:SAMPLING_INV_LUT}
\displaystyle
\begin{split}
\ddot{h}(i, \, N) & = \frac{1}{k_{i}^{\frac{N \minus 1}{2}}}  \sum_{q = 0}^{\lambda^2 \CEIL{\bar{k}^2}} \PAREN{ \ddot{S}^{\CBRAK{\lambda}} \PAREN{q + 1, \,  N} - \ddot{S}^{\CBRAK{\lambda}} \PAREN{q, \, N} } \\
& \times  L \PAREN{i - \frac{\sqrt{q}}{c T_s \lambda}, \, \alpha},  \quad  \textrm{Kernel size }  \alpha \geq 0
\end{split}
\raisetag{4ex}
\end{equation}
where $L(x, \alpha)$ is the Lanczos kernel that interpolates the forward-difference operator when $\ABS{x} \leq \alpha$, and is otherwise $0$. This construction places non-zero supports at the distance intervals containing image-sources as indicated by non-zero finite differences. Moreover, the density of the distance intervals in $k$ increases for larger area $q$, and accommodates for increasing number of image-sources at further distances. This differs from the the forward construction method, which uniformly samples over $k$ across all distances. The comparison in Fig. \ref{FIG:INVERSE_GCP} shows the differences, where the inverse construction is free of aliasing,  no-longer strictly non-negative, and reports higher echo density in the late reverberation tail that is converging towards the expected Gaussian distribution.

\subsection{Time-Frequency Controls} \label{SEC:TIME_FREQ}

GCP-ISM forward and inverse RIR constructions $\ddot{h}(i, \, N)$ in Eqs. \eqref{EQ:SAMPLING_FORWARD_LUT}, \eqref{EQ:SAMPLING_INV_LUT} support the complex reflection coefficient vectors $\VEC{\Gamma_{+}}(\omega)$, $\VEC{\Gamma_{-}}(\omega) \in \field{C}^N $. Computing $\ddot{h}(i, \, N)$ over each of $P$ uniform spaced angular frequencies $\omega_m$ between $[0, \, \pi )$ yields the time-frequency response matrix $\MAT{H} \in \field{C}^{M \times P}$ given by
\begin{equation} \label{EQ:ASSEMBLE_FREQ}
\displaystyle
\begin{split}
H(i, \, \omega) & = \ddot{h} \PAREN{i, \, N \, | \, \VEC{\Gamma_{+}}(\omega), \, \VEC{\Gamma_{-}}(\omega) }, \quad 
\omega_m = \frac{2\pi m}{P} , 
\\
h(i) & = \frac{1}{P} \sum_{n = 0}^{P \minus 1} \sum_{m = 0}^{P \minus 1}  H(i - n, \, \omega_{m}) e^{j \frac{2 \pi}{P}  m n  }.
\end{split}
\end{equation}
We can recover the RIR by taking the inverse FFT w.r.t. the rows of $\MAT{H}$ to yield a time-domain matrix. The latter's columns are then summed with sample delay equal to their column indices $n - 1$ to give the combined RIR $\VEC{h} \in \field{R}^{M}$ as expressed in Eq. \eqref{EQ:ASSEMBLE_FREQ}. The assembly therefore requires $\BIGO{M P \log P}$ compute costs.

\begin{figure}[tb]
\centerline{\includegraphics[width=0.9\columnwidth]{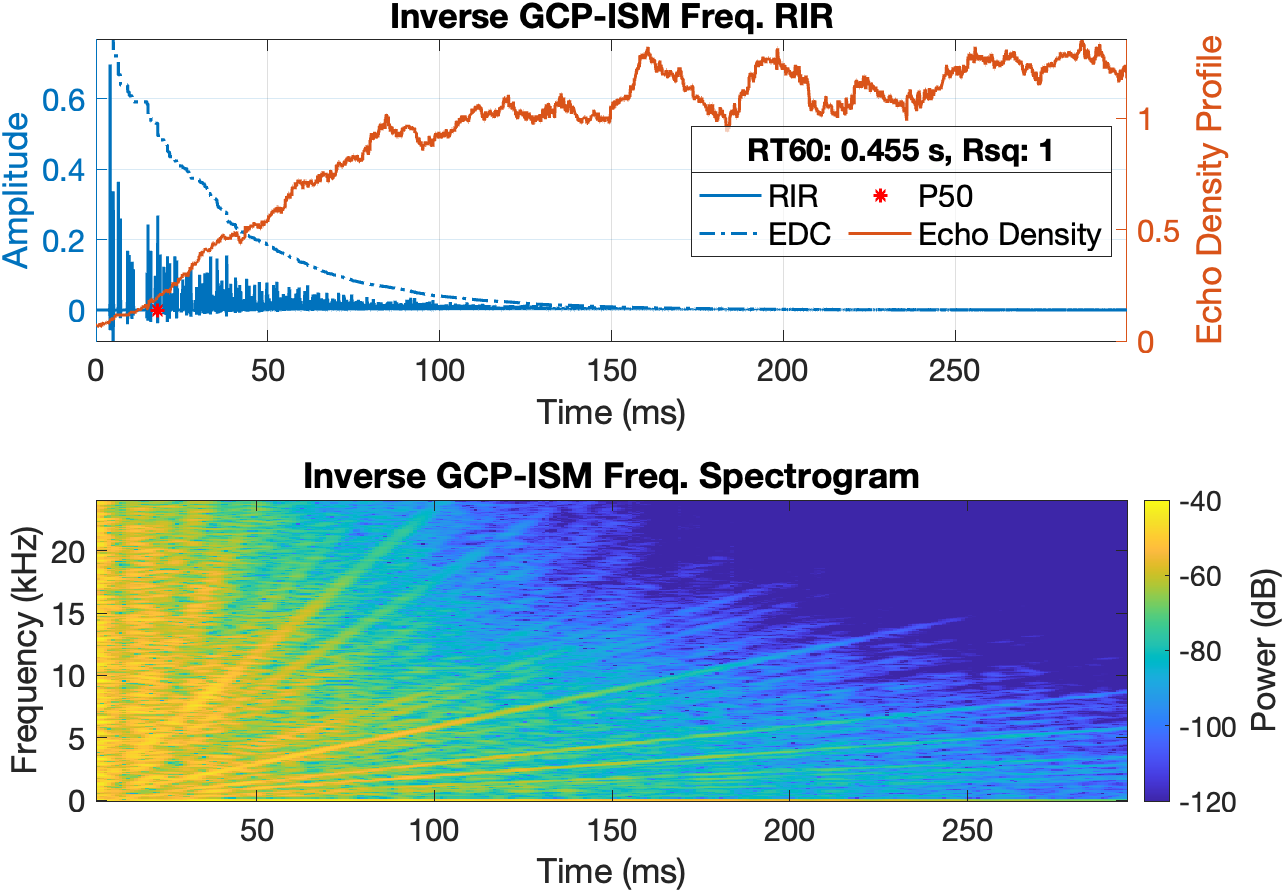}}
\caption{\label{FIG:MATERIAL_REFL}{RIR simulations with frequency-dependent reflection coefficients are assembled via Eq. \eqref{EQ:ASSEMBLE_FREQ} to yield a single RIR that reduces the high-frequency decay time as shown in the spectrogram. RIRs are generated with parameters from Fig. \ref{FIG:FORWARD_INVERSE_GCP}. We multiply the wall reflection coefficients $\VEC{\Gamma}_{+}$, $\VEC{\Gamma_{\minus}}$  by the frequency responses of a two-tap filter $\BRAK{0.8749, \, 0.1251}$ with $0$ dB at DC and $-2.5$ dB at Nyquist over $P=32$ uniform-spaced frequencies. }}
\end{figure}

For specifying $\VEC{\Gamma}(\omega)$, we define $T_{60}(\omega)$ to be the time (seconds) for $60$ dB attenuation at frequency $\omega$. The magnitude components of $\VEC{\Gamma}(\omega)$ that yield $60$ dB of attenuation at $\omega$ across axial reflections between walls along the same dimension follows
\begin{equation} \label{EQ:RT60}
\displaystyle
\begin{split}
\ABS{\VEC{\Gamma}(\omega)}_{dB} = \frac{- 60 \, \VEC{\ell} }{ c T_{60}(\omega) \sqrt{\xi}},
\end{split}
\end{equation}
where $\xi$ is the expected power of any stochastic phase-inversion process w.r.t. the sign of each reflection ($2$ if randomized, $1$ otherwise)  \cite{luo2021fast}. The phase-component  $\angle \, \VEC{\Gamma}(\omega)$ depends on the wall-material's acoustic impedance, which may induce large group-delays when added for each reflection; larger $P$ is required to prevent aliasing. Therefore, we recommend $\VEC{\Gamma}(\omega)$ to be in minimum-phase as shown in Fig. \ref{FIG:MATERIAL_REFL} whereby the reflection coefficients attenuated only the high-frequencies using a simple one-zero filter.

\section{Experiments} \label{SEC:EXP}

\subsection{RIR Generation Performance and Error Analysis}

\begin{figure}[tb]
\centerline{\includegraphics[width=0.85\columnwidth]{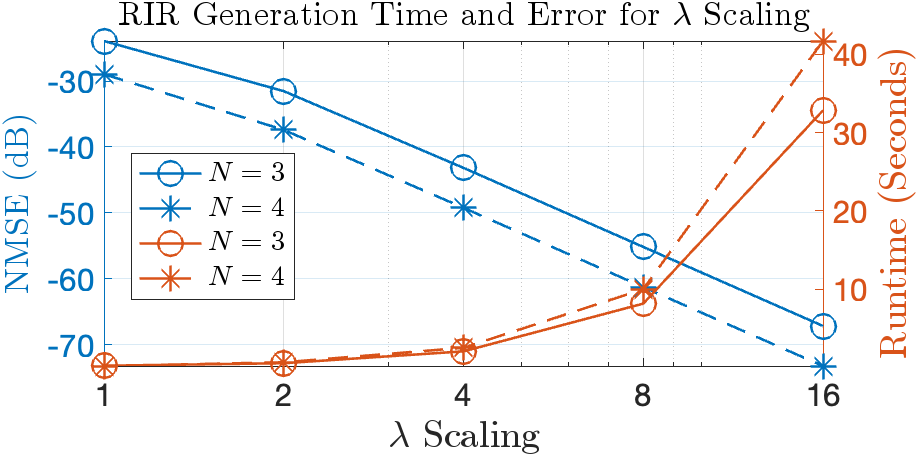}}
\caption{\label{FIG:RIR_ERROR_RUNTIME}{Doubling the scaling parameter $\lambda$ for inverse GCP-ISM  decreases the normalized mean squared error (NMSE) w.r.t. the reference ISM by $12$ dB, and more then quadruples the run-time.}}
\end{figure}

We evaluate the error to run-time trade-off of $\lambda$-scaled GCP-ISM in Eq. \eqref{EQ:SAMPLING_INV_LUT} when compared to the reference ISM RTF in Eq. \eqref{EQ:RTF}. Recall that scaling the parameters of GCP-ISM in Eq. \eqref{EQ:GCP_ISM_SCALE_EQUIV} by $\lambda$ solves for the equivalent problem with higher density of points in the volume function $\ddot{S}^{\CBRAK{\lambda}}(\lambda^2 q, N)$ as seen in Fig. \ref{FIG:MEMO_MAP}. Let us construct inverse GCP-ISM RIRs at increasing $\lambda$ using the first $N$ dimensions from parameters in Fig. \ref{FIG:INVERSE_GCP_HIGH}. We then increase the distance-equivalent time parameter to $T = 0.5$, and double the RT60\footnote{We compute EDC via Schroeder's reverse integration \cite{schroeder1965new}, estimate RT60 via linear regression of the EDC in decibels / time  between $-10$ dB and $-30$ dB, and report the R-squared coefficient of determination.} time by raising all reflection coefficients to $\Gamma_{+ n}^{1/2}$, $\Gamma_{\minus n}^{1/2}$. The error to run-time trade-offs are shown in Fig. \ref{FIG:RIR_ERROR_RUNTIME} whereby each doubling of $\lambda$ reduces the error by $12 $ dB at more than quadruple the run-time  due to  $\BIGO{N (k \lambda)^2 \log k \lambda}$ compute costs.

\begin{figure}[tb]
\centerline{\includegraphics[width=0.8\columnwidth]{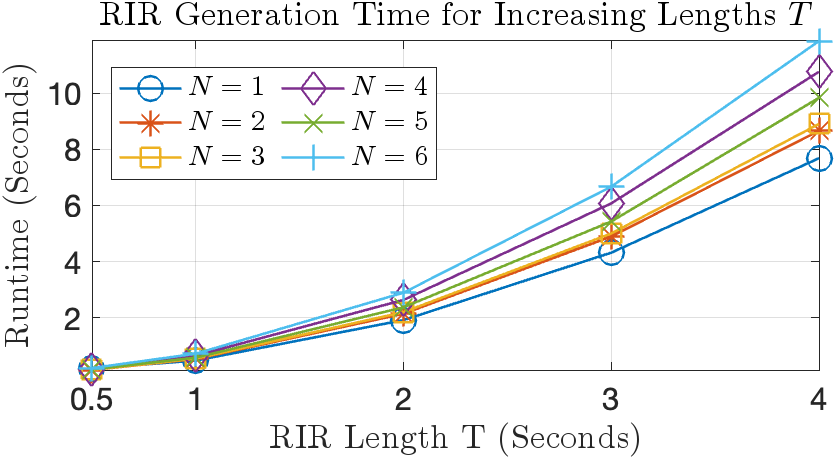}}
\caption{\label{FIG:RIR_RT60_CTR}{Doubling the RIR length $T$ more than quadruples the run-time costs, regardless of dimension $N$.}}
\end{figure}

\begin{figure}[tb]
\centerline{\includegraphics[width=0.95\columnwidth]{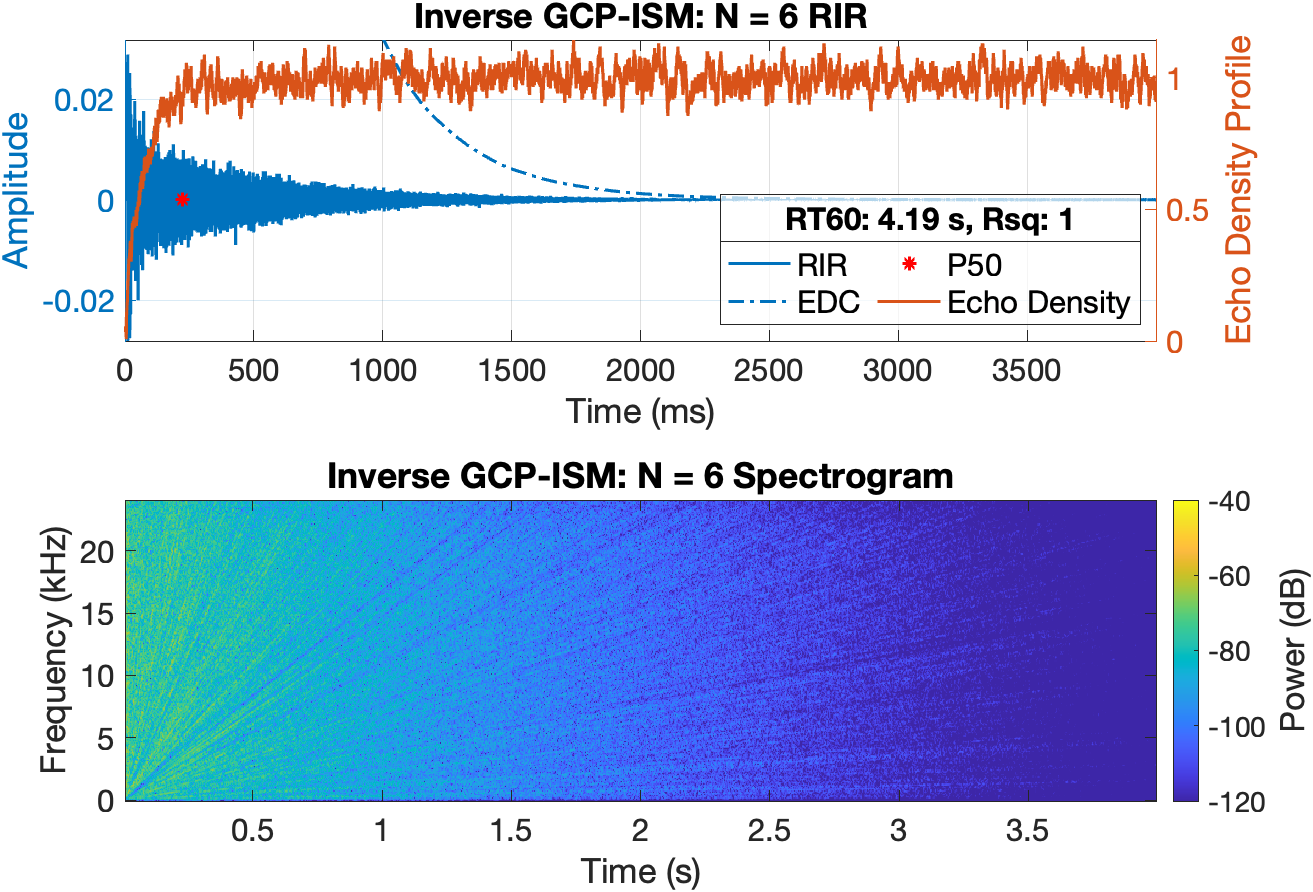}}
\caption{\label{FIG:RIR_RT60_CTR_SAMPLE}{The sample RIR in $N=6$ dimensions consists of reflection coefficients $\VEC{\Gamma_{+}}$, $\VEC{\Gamma_{\minus}}$ designed for $T_{60} \sim 4$ across all axial reflections. The mixing time is fast due to the rapid growth of the echo density in high dimensions; echo density converges to $1$ around the time of the $50^{th}$ energy percentile. The late-tail contains half the energy of the RIR, sounds acoustically smooth without flutters, and presents some time-evolving colorations.}}
\end{figure}

We now consider increasing both the range of RIR lengths to $T \leq 4$ and dimensions $N \leq 6$ for evaluating inverse GCP-ISM RIR generation times. Parameters are taken from Fig. \ref{FIG:INVERSE_GCP_HIGH} and the reflection coefficients are set to achieve broadband $T_{60} \sim T$ by setting $\xi=4$ in Eq. \eqref{EQ:RT60}. The run-time performance shown in Fig. \ref{FIG:RIR_RT60_CTR} tracks with the $\lambda$-scaling trade-offs in Fig. \ref{FIG:RIR_ERROR_RUNTIME} as doubling the former's RIR length also quadruples the number of points memoized in the volume function.
We also note that doubling the number of dimensions did not double the run-times due to costs of the inverse RIR construction in Eq. \eqref{EQ:SAMPLING_INV_LUT}. The sample RIR of both the longest $T_{60}=4$ and highest dimension $N=6$ is shown in Fig. \ref{FIG:RIR_RT60_CTR_SAMPLE}.

\subsection{High-Dimensional Echo Density Study}

In GCP-ISM, we expect the number of image-sources within a $k$-radius ball of $N$ dimensional orthotopes to be proportional to the ball's volume or $k^N$, and the density on the ball's surface to be of $k^{N \minus 1}$. The latter derives the distance intensity attenuation component $A(\VEC{v})$ in Eq. \eqref{EQ:RTF} after taking the square root. However, the distribution of image-source intensities on the ball's surface differs from that of GCP, due to the asymmetries in the orthotope size, reflection coefficients, and source-receiver coordinates. The echo density profile $\eta(t)$ in \cite{abel2006simple} is one such useful measure where $\eta(t)$ gives the fraction of samples that lie outside one standard deviation within the interval centered on $t$ relative to that of a normal distribution; reverberation tails typically converge to $\eta(t) = 1$.

\begin{figure*}[tb!]
\centering 
    \subfloat[$N=4$  \label{FIG:INVERSE_GCP_N_4}]{%
        \includegraphics[width=0.33\textwidth]{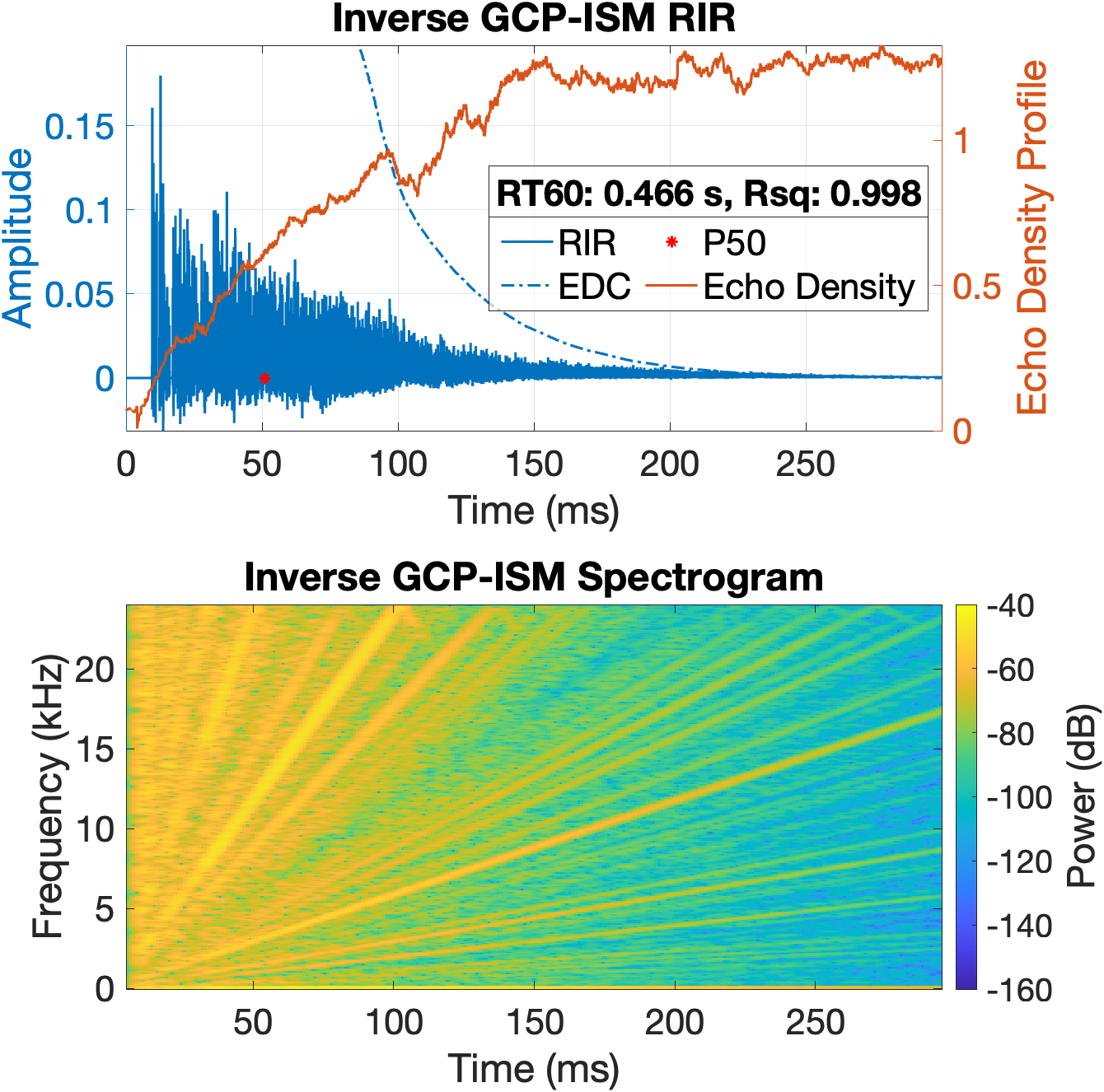}%
    }
    \subfloat[$N=5$  \label{FIG:INVERSE_GCP_N_5}]{%
        \includegraphics[width=0.33\textwidth]{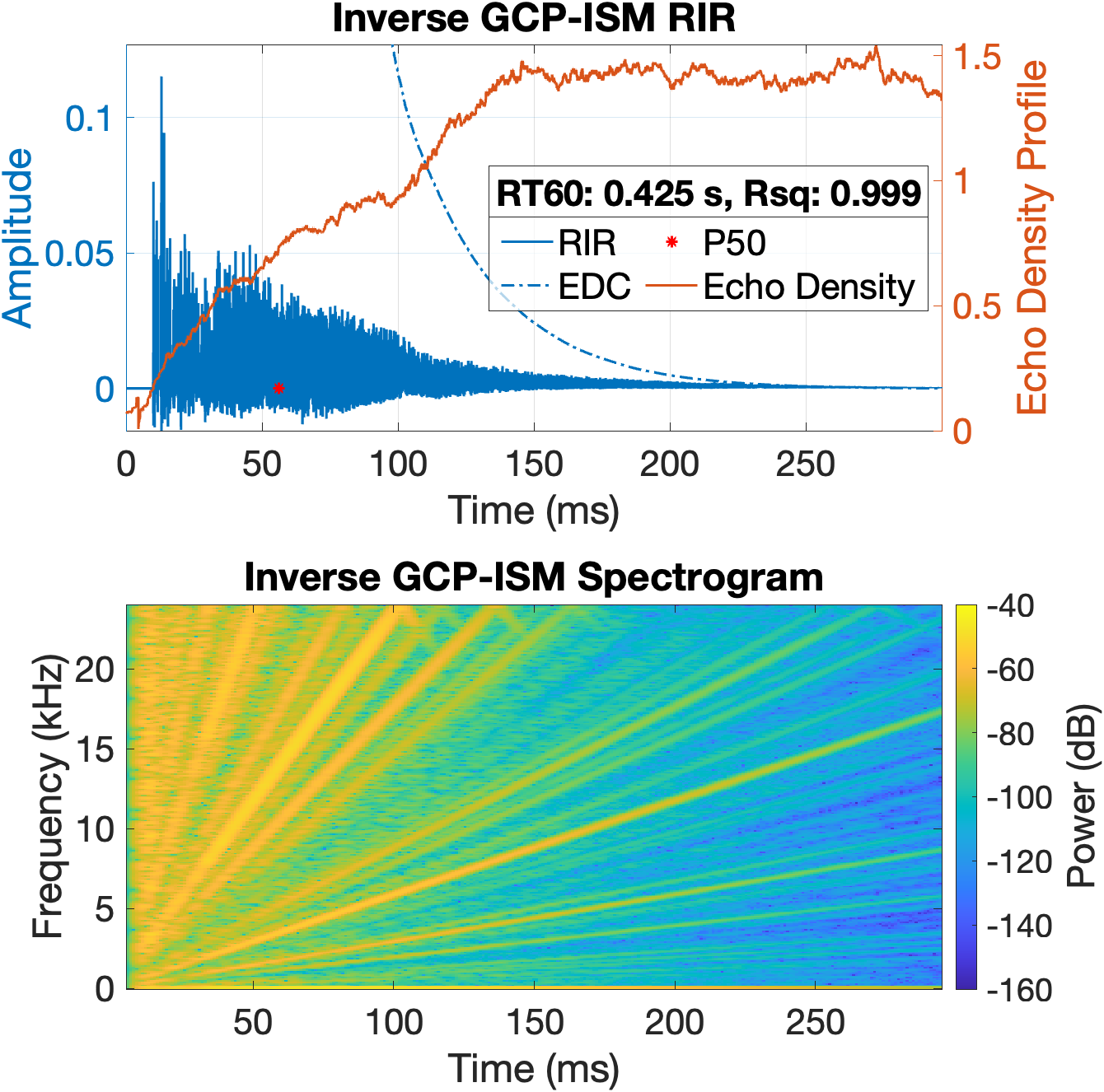}%
    }
    \subfloat[$N=6$  \label{FIG:INVERSE_GCP_N_6}]{%
        \includegraphics[width=0.33\textwidth]{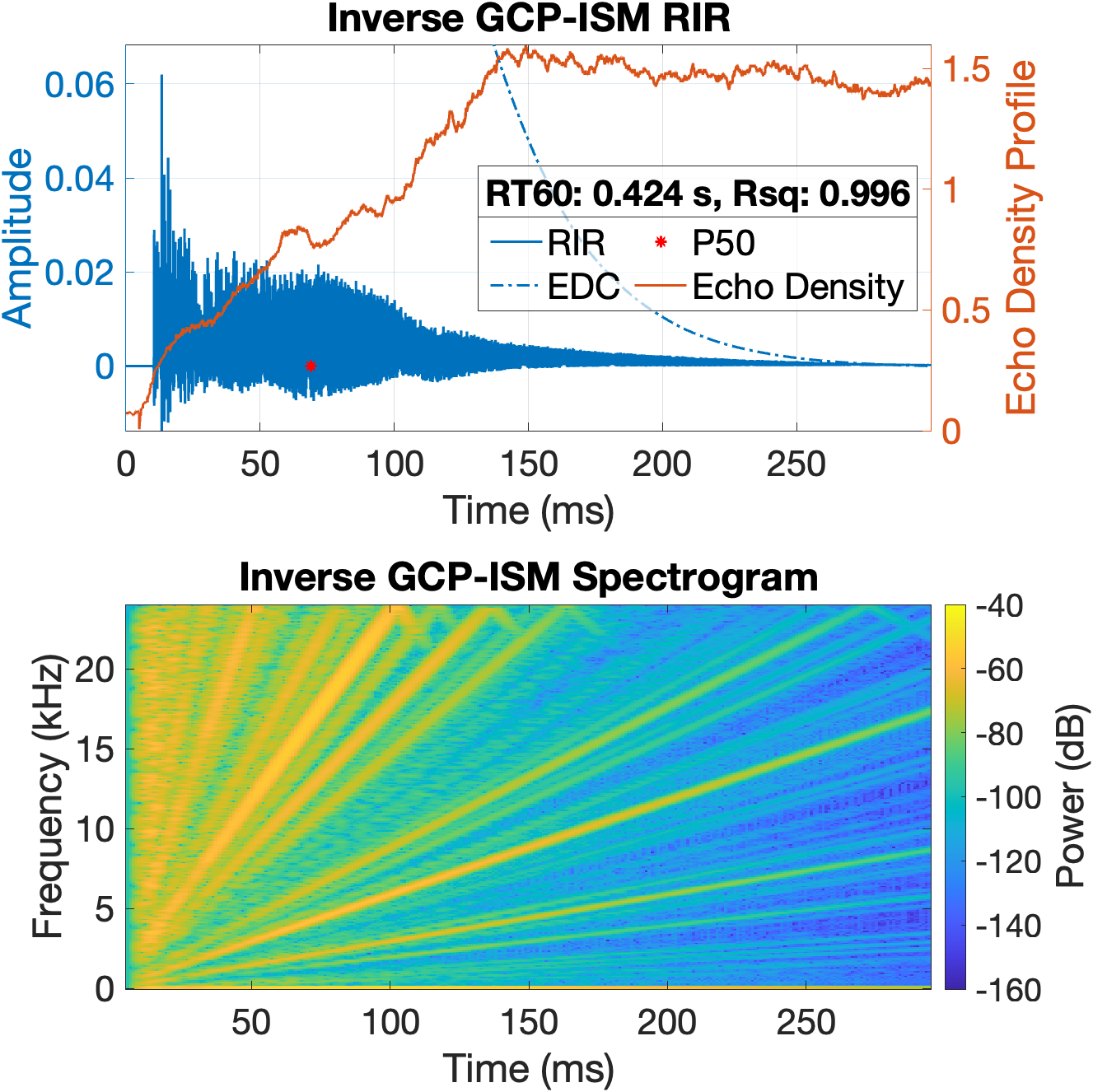}%
    }
      \caption{\label{FIG:INVERSE_GCP_HIGH}  Echo density of higher-dimensional ($N > 3$) GCP-ISM RIRs converges to $1.5$ (higher density in distribution's tail than Gaussian) in the simulation. RIRs are generated from the first $N$ dimensions of the following parameters: $\VEC{s} = \BRAK{1, 0, 1, 0,      1,       3}$, $\VEC{r} = \BRAK{2, 1, 1, 3,      2,       2}$, $\VEC{\ell} = \BRAK{5, 4, 3, 7,      6,       8}$, $\VEC{\Gamma_{+}} = \BRAK{0.93, 0.8,  0.9, 0.93,   0.77,   0.82}$, $\VEC{\Gamma_{\minus}} = \BRAK{0.72, 0.78, 0.93, 0.67,   0.52,   0.7}$, $\lambda = 1$, and $T = 0.3$.}  
\end{figure*}

\begin{figure*}[tb!]
\centering 
    \subfloat[Reflection coeffs. $-\VEC{\Gamma}_{+}$, $-\VEC{\Gamma_{-}}$  \label{FIG:INVERSE_GCP_N_6_REFL_POSNEG}]{%
        \includegraphics[width=0.33\textwidth]{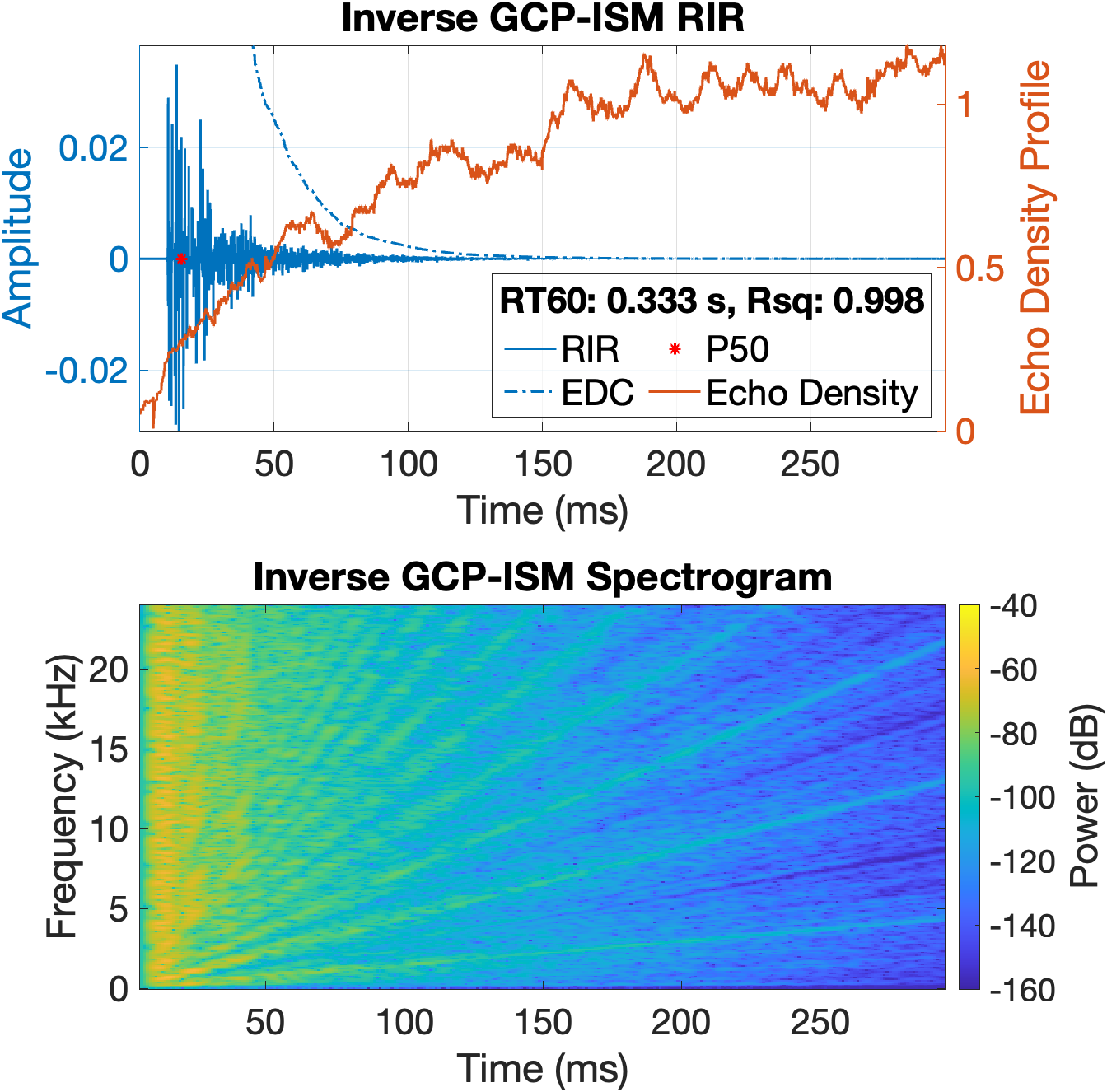}%
    }
    \subfloat[Reflection coeffs. $-\VEC{\Gamma}_{+}$, $+\VEC{\Gamma_{-}}$  \label{FIG:INVERSE_GCP_N_6_REFL_POS}]{%
        \includegraphics[width=0.33\textwidth]{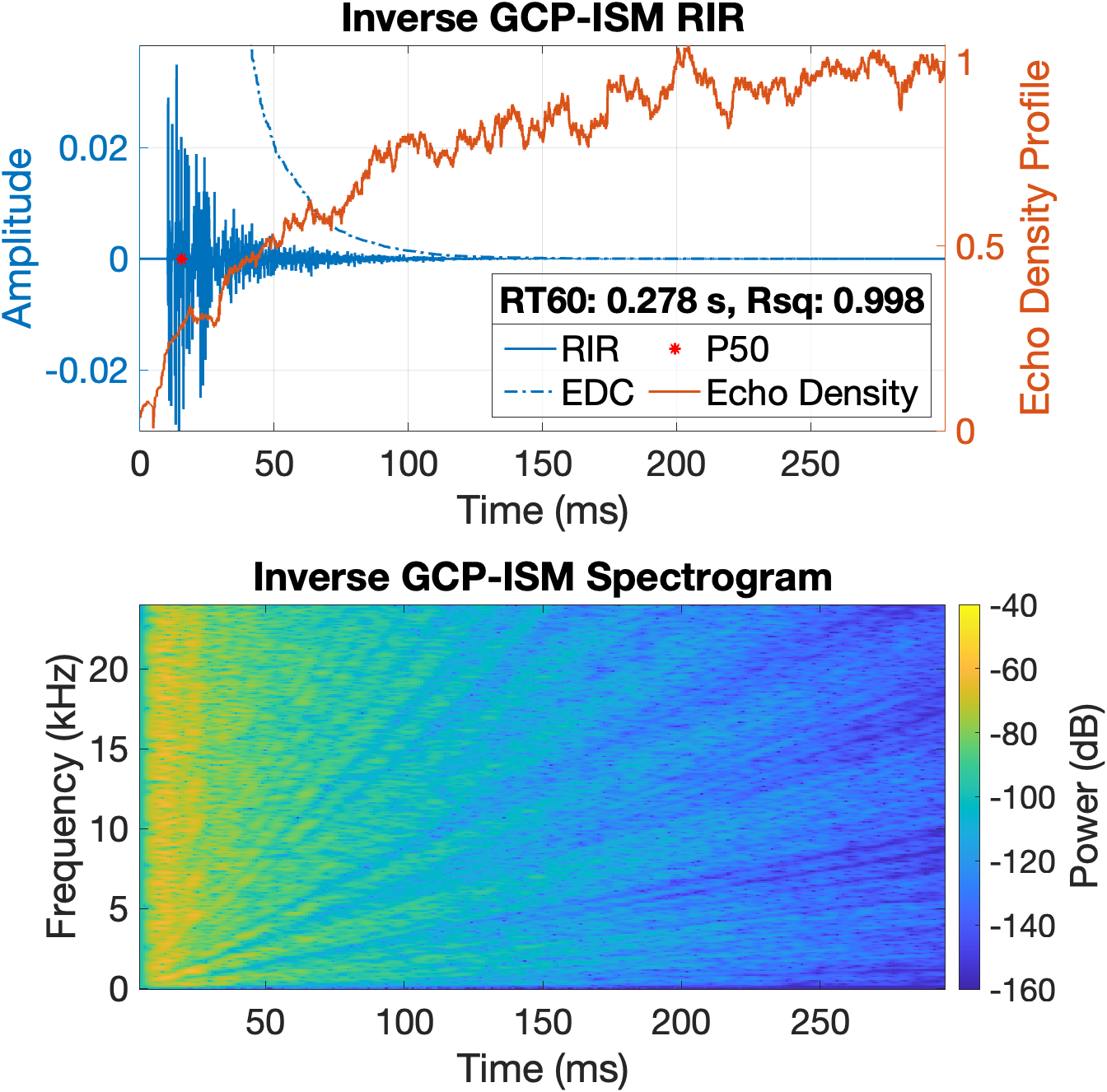}%
    }
    \subfloat[Reflection coeffs.  $(-1)^{n} \Gamma_{+n}$, $(-1)^{n + 1} \Gamma_{\minus n}$  \label{FIG:INVERSE_GCP_N_6_REFL_ALT}]{%
        \includegraphics[width=0.33\textwidth]{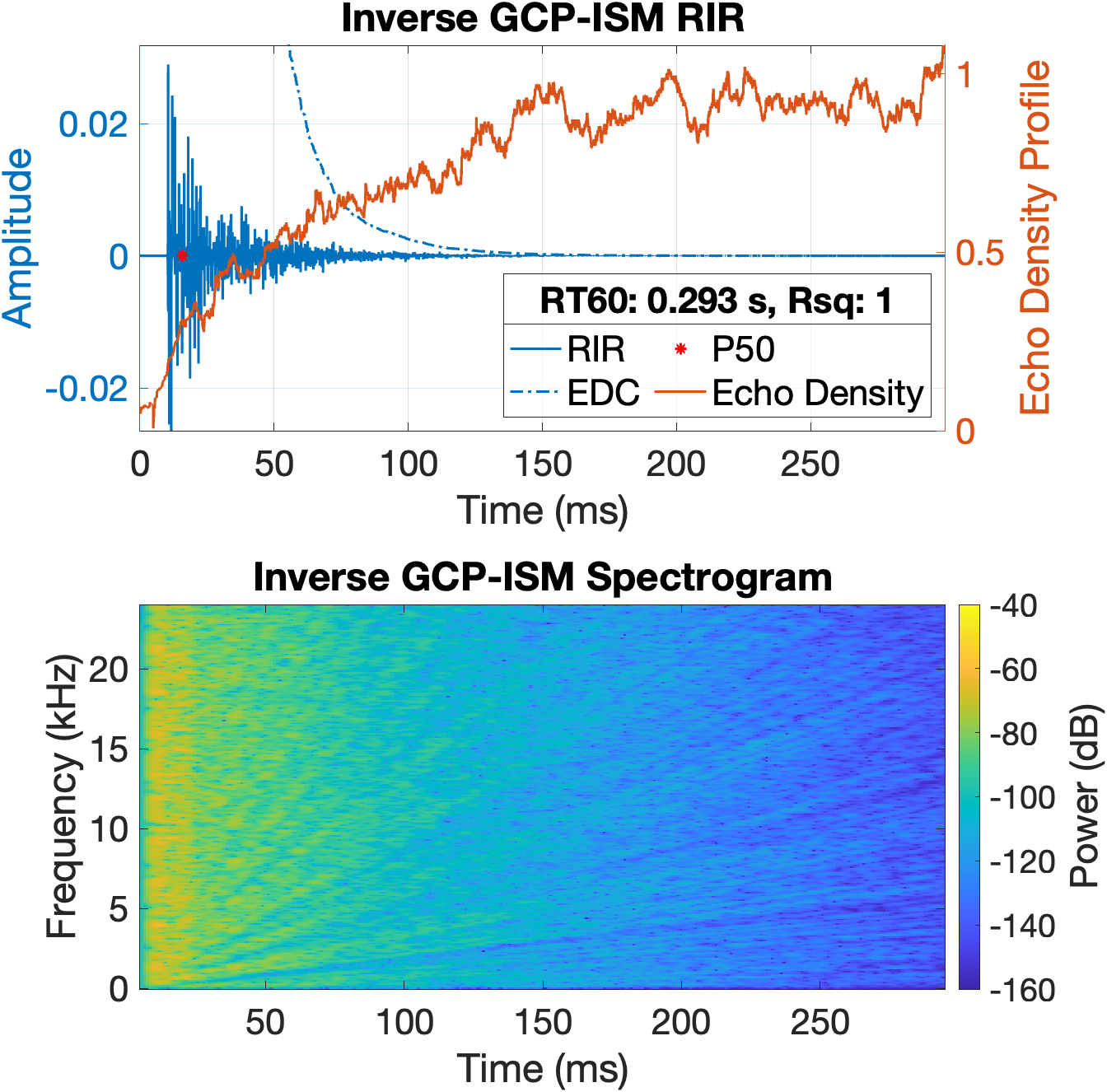}%
    }
      \caption{\label{FIG:INVERSE_GCP_HIGH_REFL} Echo density converges to $1$ in RIR simulations with non-resistive acoustic impedances; real reflection coefficients belonging to all, positive-facing, or alternating facing walls are phase-flipped. RIRs are generated for $N=6$ dimensions with parameters in Fig. \ref{FIG:INVERSE_GCP_HIGH}. }  
\end{figure*}

Indeed, earlier simulations in Fig. \ref{FIG:FORWARD_INVERSE_GCP} show RIR tail distributions approaching Gaussian. However, generating RIRs in higher-dimension orthotopes ($4 \leq N \leq 6$) via the GCP-ISM inverse construction method of Eq. \eqref{EQ:SAMPLING_INV_LUT} shows higher echo densities $\eta(t) > 1$ in excess of Gaussian within the late reverb distribution's tail as seen in Fig. \ref{FIG:INVERSE_GCP_HIGH}. Several factors contribute to the divergence: The greater surface area of the GCP-ISM $k$-radius ball for larger dimensions $N$ accelerates the build-up and total number of reflections for increasing $k$. The reflection coefficients in the simulation are strictly in-phase and resistive (positive real impedance), resulting in the dominance of specific frequency modes in the reverberation tail after $150$ ms. Non-resistive acoustic materials can have negative-real reflection coefficients, which when mixed in simulation, generate reverberation tails with the expected Gaussian echo density as shown in Fig. \ref{FIG:INVERSE_GCP_HIGH_REFL}; visible modes in the spectrogram are scattered. Note that real-world acoustic impedance of wall materials are complex due to varying porosity and non-zero reactance (imaginary components), which we support in our model. 

\section{Conclusions} \label{SEC:CONCLUSION}

We presented a fast computational method for high-dimensional ISM in integer coordinates $\field{Z}^N$ that reduces its computational complexity class to $\BIGO{N k^2 \log k}$. Our method extended the classic 2D lattice counting GCP into higher dimensions by expressing sub-solutions across dimensions via recurrence relations and fast geometric convolution. We then extended GCP to ISM, which included coordinate translation, scaling, and weighting to model varied source, receiver locations, and room reflections. RIRs were realized via forward and inverse finite difference methods of a GCP-ISM volume function, which can be made more accurate via coordinate scaling. Later experiments showed the trade-offs between error-reduction and run-time per doubling of coordinate dimensions. We supported frequency-dependent reflections via complex reflection coefficients and RIR assembly via the inverse Fourier transform and time-shifted summation. Experiments showed that RIR in high-dimensions with only positive-real reflection coefficients had excessive echo density, which was addressed via the introduction of phase-inverted reflection coefficients. Long duration and high-dimension RIRs were shown with metrics for posterity. 


 
\bibliographystyle{IEEEtranDAFx}
\bibliography{DAFx26_tmpl} 

@inproceedings{abel2006simple,
  title={A simple, robust measure of reverberation echo density},
  author={Abel, Jonathan S and Huang, Patty},
  booktitle={Audio Engineering Society Convention 121},
  year={2006},
  organization={Audio Engineering Society}
}

@inproceedings{jot1997analysis,
  title={Analysis and synthesis of room reverberation based on a statistical time-frequency model},
  author={Jot, Jean-Marc and Cerveau, Laurent and Warusfel, Olivier},
  booktitle={AES: Convention of the Audio Engineering Society},
  year={1997}
}

@article{georganti2008analysis,
  title={Analysis of room transfer function and reverberant signal statistics: Abstract of paper},
  author={Georganti, Eleftheria and Mourjopoulos, John and Jacobsen, Finn},
  journal={Acoustical Society of America. Journal},
  volume={123},
  number={5},
  pages={3761--3761},
  year={2008},
  publisher={Acoustical Society of America}
}

@article{schlecht2016lossless,
  title={On lossless feedback delay networks},
  author={Schlecht, Sebastian J and Habets, Emanu{\"e}l AP},
  journal={IEEE Transactions on Signal Processing},
  volume={65},
  number={6},
  pages={1554--1564},
  year={2016},
  publisher={IEEE}
}

@article{gerzon1971synthetic,
  title={Synthetic stereo reverberation: Part one},
  author={Gerzon, Michael A},
  journal={Studio Sound},
  volume={13},
  number={12},
  pages={632--635},
  year={1971}
}

@inproceedings{jot1991digital,
  title={Digital delay networks for designing artificial reverberators},
  author={Jot, Jean-Marc and Chaigne, Antoine},
  booktitle={Audio Engineering Society Convention 90},
  year={1991},
  organization={Audio Engineering Society}
}

@article{allen1979image,
  title={Image method for efficiently simulating small-room acoustics},
  author={Allen, Jont B and Berkley, David A},
  journal={The Journal of the Acoustical Society of America},
  volume={65},
  number={4},
  pages={943--950},
  year={1979},
  publisher={Acoustical Society of America}
}

@article{aretz2014application,
  title={Application of the mirror source method for low frequency sound prediction in rectangular rooms},
  author={Aretz, Marc and Dietrich, Pascal and Vorl{\"a}nder, Michael},
  journal={Acta Acustica united with Acustica},
  volume={100},
  number={2},
  pages={306--319},
  year={2014},
  publisher={S HIRZEL VERLAG POSTFACH 10 10 61, D-70 009 STUTTGART, GERMANY}
}

@inproceedings{scheibler2018pyroomacoustics,
  title={Pyroomacoustics: A python package for audio room simulation and array processing algorithms},
  author={Scheibler, Robin and Bezzam, Eric and Dokmani{\'c}, Ivan},
  booktitle={2018 IEEE international conference on acoustics, speech and signal processing (ICASSP)},
  pages={351--355},
  year={2018},
  organization={IEEE}
}

@article{borish1984extension,
  title={Extension of the image model to arbitrary polyhedra},
  author={Borish, Jeffrey},
  journal={The Journal of the Acoustical Society of America},
  volume={75},
  number={6},
  pages={1827--1836},
  year={1984},
  publisher={Acoustical Society of America}
}

@article{rindel2000use,
  title={The use of computer modeling in room acoustics},
  author={Rindel, Jens Holger},
  journal={Journal of vibroengineering},
  volume={3},
  number={4},
   IGNOREpages={219--224},
  year={2000},
  publisher={EXTRICA}
}

@article{habets2006room,
  title={Room impulse response generator},
  author={Habets, Emanuel AP},
  journal={Technische Universiteit Eindhoven, Tech. Rep},
  volume={2},
  number={2.4},
   IGNOREpages={1},
  year={2006}
}

@inproceedings{schissler2011gsound,
  title={Gsound: Interactive sound propagation for games},
  author={Schissler, Carl and Manocha, Dinesh},
  booktitle={Audio Engineering Society Conference: 41st International Conference: Audio for Games. Audio Engineering Society},
  volume={6},
  year={2011}
}

@article{lehmann2009diffuse,
  title={Diffuse reverberation model for efficient image-source simulation of room impulse responses},
  author={Lehmann, Eric A and Johansson, Anders M},
  journal={IEEE Transactions on Audio, Speech, and Language Processing},
  volume={18},
  number={6},
  pages={1429--1439},
  year={2009},
  publisher={IEEE}
}

@article{heinz1993binaural,
  title={Binaural room simulation based on an image source model with addition of statistical methods to include the diffuse sound scattering of walls and to predict the reverberant tail},
  author={Heinz, Renate},
  journal={Applied Acoustics},
  volume={38},
  number={2-4},
   IGNOREpages={145--159},
  year={1993},
  publisher={Elsevier}
}

@article{ali2025source,
  title={Source-time dominant modeling of the Doppler shift for the auralization of moving sources},
  author={Ali, Randall and Christian, Andrew},
  journal={Acta Acustica},
  volume={9},
  pages={1},
  year={2025},
  publisher={EDP Sciences}
}

@inproceedings{mcgovern2011image,
  title={The image-source reverberation model in an N-dimensional space},
  author={McGovern, Stephen},
  booktitle={14th International conference on digital audio effects},
  pages={11--18},
  year={2011}
}

@book{kuttruff2016room_normalmodes,
  title={Room acoustics},
  author={Kuttruff, Heinrich},
  year={2000},
  publisher={Crc Press},
  Edition = {Fourth},
  pages={65}
}

@phdthesis{thomas2017wayverb,
  title={Wayverb: A graphical tool for hybrid room acoustics simulation},
  author={Thomas, Matthew Reuben},
  year={2017},
  school={University of Huddersfield}
}

@article{xu2024simulating,
  title={Simulating room transfer functions between transducers mounted on audio devices using a modified image source method},
  author={Xu, Zeyu and Herzog, Adrian and Lodermeyer, Alexander and Habets, Emanu{\"e}l AP and Prinn, Albert G},
  journal={The Journal of the Acoustical Society of America},
  volume={155},
  number={1},
  pages={343--357},
  year={2024},
  publisher={AIP Publishing}
}

@article{schroeder1965new,
  title={New method of measuring reverberation time},
  author={Schroeder, Manfred R},
  journal={The Journal of the Acoustical Society of America},
  volume={37},
  number={3},
  pages={409--412},
  year={1965},
  publisher={Acoustical Society of America}
}

@inproceedings{luo2021fast,
  title={Fast source-room-receiver acoustics modeling},
  author={Luo, Yuancheng and Kim, Wontak},
  booktitle={2020 28th European Signal Processing Conference (EUSIPCO)},
  pages={51--55},
  year={2021},
  organization={IEEE}
}

@article{samarasinghe2018spherical,
  title={Spherical harmonics based generalized image source method for simulating room acoustics},
  author={Samarasinghe, Prasanga N and Abhayapala, Thushara D and Lu, Yan and Chen, Hanchi and Dickins, Glenn},
  journal={The Journal of the Acoustical Society of America},
  volume={144},
  number={3},
   IGNOREpages={1381--1391},
  year={2018},
  publisher={AIP Publishing}
}

@book{sabine2015collected,
  title={Collected papers on acoustics},
  author={Sabine, Wallace Clement},
  year={2015},
  publisher={Courier Dover Publications}
}

@article{eyring1930reverberation,
  title={Reverberation time in “dead” rooms},
  author={Eyring, Carl F},
  journal={The Journal of the Acoustical Society of America},
  volume={1},
  number={2A\_Supplement},
  pages={168--168},
  year={1930},
  publisher={AIP Publishing}
}

@article{lehmann2008prediction,
  title={Prediction of energy decay in room impulse responses simulated with an image-source model},
  author={Lehmann, Eric A and Johansson, Anders M},
  journal={The Journal of the Acoustical Society of America},
  volume={124},
  number={1},
  pages={269--277},
  year={2008},
  publisher={AIP Publishing}
}

@inproceedings{kelloniemi2005artificial,
  title={Artificial reverberation using a hyper-dimensional {FDTD} mesh},
  author={Kelloniemi, Antti and V{\"a}lim{\"a}ki, Vesa and Huang, Patty and Savioja, Lauri},
  booktitle={13th European Signal Processing Conference},
  pages={1--4},
  year={2005},
  organization={IEEE}
}


\section{Appendix}

To compute $\bar{g}(x, y)  = \argmax{m \in \field{Z} } m  \quad \textrm{s.t.} \quad  m + (-1)^m y \leq x$, find the maximum even and odd valued $m$ given by
\begin{equation} \label{EQ:APP:MAX_INT}
\displaystyle
\begin{split}
m & = 2k \,\,\, \Rightarrow \,\,\,
 2k + y \leq x \,\,\,  \Rightarrow \,\,\,
 k \leq \frac{x - y}{2}, \\
m & = 2k + 1 \,\,\, \Rightarrow \,\,\,
 2k + 1 - y \leq x \,\,\,  \Rightarrow \,\,\,
 k \leq \frac{x + y - 1}{2}, \\
\Rightarrow & \quad  \bar{g}(x, y)   = \max \CBRAK{ 2 \FLOOR{\frac{x - y}{2}}, \,  2 \FLOOR{\frac{x + y - 1}{2}} + 1}.
\end{split}
\raisetag{15.5ex}
\end{equation}

To compute $\underline{g}(x, y)  = \argmin{m \in \field{Z} } m  \quad \textrm{s.t.} \quad  m + (-1)^m y \geq x$, find the minimum even and odd valued $m$ given by
\begin{equation} \label{EQ:APP:MIN_INT}
\displaystyle
\begin{split}
m & = 2k \,\,\, \Rightarrow \,\,\,
 2k + y \geq x \,\,\,  \Rightarrow \,\,\,
 k \geq \frac{x - y}{2}, \\
m & = 2k + 1 \,\,\, \Rightarrow \,\,\,
 2k + 1 - y \geq x \,\,\,  \Rightarrow \,\,\,
 k \geq \frac{x + y - 1}{2}, \\
\Rightarrow & \quad  \underline{g}(x, y)   = \min \CBRAK{ 2 \CEIL{\frac{x - y}{2}}, \,  2 \CEIL{\frac{x + y - 1}{2}} + 1}.
\end{split}
\raisetag{15.5ex}
\end{equation}


\end{document}